\documentclass[12pt,leqno]{article}
\usepackage{amsthm,amsfonts,amssymb,amsmath}
\usepackage{epsfig}
\usepackage{graphicx}
\numberwithin{equation}{section}
\usepackage[thinlines]{easybmat}
\usepackage{epstopdf} 
\usepackage{showkeys}

\usepackage{pstricks}
\newcommand\R{\mathbb R}

\newcommand\br{\begin{remark}}
\newcommand\er{\end{remark}}
\newcommand\bp{\begin{pmatrix}}
\newcommand\ep{\end{pmatrix}}
\newcommand\be{\begin{equation}}
\newcommand\ee{\end{equation}}
\newcommand\ba{\begin{equation}\begin{aligned}}
\newcommand\ea{\end{aligned}\end{equation}}

\newcommand{\bap}{\begin{app}}
\newcommand{\eap}{\end{app}}
\newcommand{\begs}{\begin{exams}}
\newcommand{\eegs}{\end{exams}}
\newcommand{\beg}{\begin{example}}
\newcommand{\eeg}{\end{exaplem}}
\newcommand{\bpr}{\begin{proposition}}
\newcommand{\epr}{\end{proposition}}
\newcommand{\bt}{\begin{theorem}}
\newcommand{\et}{\end{theorem}}
\newcommand{\bc}{\begin{corollary}}
\newcommand{\ec}{\end{corollary}}
\newcommand{\bl}{\begin{lemma}}
\newcommand{\el}{\end{lemma}}
\newcommand{\brs}{\begin{remarks}}
\newcommand{\ers}{\end{remarks}}

\newtheorem{theorem}{Theorem}[section]
\newtheorem{proposition}[theorem]{Proposition}
\newtheorem{corollary}[theorem]{Corollary}
\newtheorem{lemma}[theorem]{Lemma}

\newtheorem{example}[theorem]{Example}
\newtheorem{remark}[theorem]{Remark}

\title{Thin power law film flow down an inclined plane: consistent shallow water models and\\ stability under large scale perturbations}

\author{\sc \small
Pascal Noble\thanks{Universit\'e de Lyon, Universit\'e Lyon I, Institut Camille Jordan, UMR CNRS 5208, 43 bd du 11 novembre 1918, F - 69622 Villeurbanne Cedex, France; noble@math.univ-lyon1.fr:
Research of P.N. is partially supported by the French ANR Project no.
ANR-09-JCJC-0103-01.}
~~~
Jean-Paul Vila\thanks{Institut de MathŽmatiques de Toulouse, UMR CNRS 5219, INSA de Toulouse, 135 avenue de Rangueil, 31077 Toulouse Cedex 4, France}
}

\begin{document}
\date{\today}
\maketitle

\begin{abstract}
The aim of this paper is to provide a better understanding of the derivation of two moments approximate models of shallow water type for thin power law films down an incline. We limit ourselves to the case of laminar flow for which the boundary layer issued from the interaction of the flow with the bottom surface has an influence all over the transverse direction to the flow. In this case the concept itself of thin film and its relation with long wave asymptotic leads naturally to flow conditions around a uniform free surface Poiseuille flow. 

This Poiseuille flow exhibits a divergence of the apparent viscosity at the free surface which, in turn, introduces a singularity in the formulation of the Orr-Sommerfeld equations and in the derivation of shallow water models. We remove this singularity by inverting the relation between the fluid strain and the deformation tensor and by keeping the fluid strain as a variable. No regularization procedure is needed here, in contrast to \cite{RQ}.

In order to derive the shallow water equations, we compute a second order asymptotic expansion of the fluid velocity field and fluid strain in the shallow water regime through a nonlinear iterative scheme. We show that a linearized version of this scheme naturally provides an expansion of the eigenfunctions of the Orr-Sommerfeld equations in the small wavenumber/small frequency regime. The shallow water models account for the streamwise diffusion of momentum which is important to describe accurately the full dynamic of the thin film flows when instabilities like roll-waves arise.
\end{abstract}

\section{Introduction}

Mathematical models and numerical simulations for the flow of a relatively thin layer of Non-Newtonian fluids over an inclined surface and under the effect of gravity have important applications in natural processes such as mud flows in mountainous areas or submarine continental slopes, debris avalanches: it is of interest to be able to predict the characteristics (velocity, pressure, runout extent) of such mud flows for geophysical and engineering purposes.

Herein, we consider the simple situation of a thin liquid layer over an inclined plane. This is not a restriction if, instead, one considers an Òarbitrary topographyÓ provided that the curvature is asymptotically small and slope never vanishes: in this case, we just substitute the constant slope by local slope of the bottom (see the discussion in \cite{BCNV}). Moreover, since instabilities in such flows are primarily two dimensional, we will assume that the flow is two dimensional. Herein, we consider an incompressible and Non-Newtonian fluid, modeled by the Ostwald-de Wael (also known as the Norton-Hoff law in metallurgy) constitutive law
\begin{equation}\label{pl_intro}
\displaystyle
\tau=\mu_p\|D({\bf{u}})\|^{n-1}D(\bf{u}).
\end{equation}
\noindent
where $\tau$ is the strain rate of the fluid, $\bf{u}$ its velocity and $D(\bf{u})=\nabla \bf{u}+\nabla\bf{u}^T$ the deformation rate. The norm on tensors $\|A\|^2={\rm Tr}(AA^T)$ is the classical Frobenius norm. The case $n<1$ corresponds to shear thinning fluids (like ice) whereas $n>1$ corresponds to shear thickening fluids (like some polymeric fluids).\\

Given the fact that the two dimensional Cauchy Momentum equations are difficult to treat both analytically and numerically, in particular if the boundary is free, it is important to obtain reduced models that are able to capture the relevant features, but are mathematically more manageable. Modeling the motion of a thin films down an inclined plane leads to a hierarchy of models. The first approximation level are the lubrication models: in the shallow water regime and if inertia is small (small Reynold number), the fluid velocity and pressure are determined by the local fluid height and its derivatives. In this case one obtains a model for the local fluid height in the form $\displaystyle \partial_t h+\partial_x G(h,\partial_x h,É,\partial_x^m h)=0$ where $G$ involves various algebraic powers and differentiation orders of $h$. This type of equations is obtained by means of asymptotic expansions in power of $\varepsilon$, usually called film parameter or aspect ratio between the width of the layer and the typical length of the phenomena in the streamwise direction. The simplest models obtained for Newtonian fluids are the Benney's \cite{Be} and generalized Kuramoto-Sivashinsky equations. The simplification brought by this reduction has permitted a first study of the nonlinear development of waves using dynamical system theory \cite{PMP}. In the case of Non-Newtonian fluids described by a power-law fluids ($0 < n < 1$ for mud flows or ice sheets dynamic),  Ng and Mei [14] obtained a  Benney type equation and found a stability criterion for Poiseuille flows under large wavelength perturbations. However, Benney's equation exhibits finite time singularities and is only valid for small amplitude waves.\\

In order to obtain models valid for thicker flows and larger Reynolds number, one considers shallow water type flows: a specific shape for the velocity profile is assumed together with a hydrostatic field assumption. Averaging the streamwise momentum equation and the divergence free condition, one obtains an evolution system for the local fluid height $h$ and the local streamwise flow rate $q$. In the Newtonian case, a first model, also called integral boundary layer model was derived by Shkadov \cite{S}, by assuming that the flow remains close to the Poiseuille flow. However this model is not consistent at order one with respect to $\varepsilon$. This results in an error on the prediction of the critical Reynolds number above which constant states are unstable under large wavelength perturbations, an information that is important to predict the onset of roll-waves. More recently, Ruyer-Quil and Manneville \cite{RQ1,RQ2} corrected this shallow water model by using a Galerkin type approach and even derived consistent second order models by introducing a third variable $\tau$ that measures the departure of the wall shear from the shear predicted by the semi parabolic velocity profile (which is precisely the quantity that has to be determined in order to find consistent models).
    
There are many attempts found in the literature to derive a shallow water model for a thin layer of power-law fluid down an inclined plane, all based on the Shkadov approach and thus inconsistent \cite{NM,BHS,DM1,DM2,HCWL}. This is all the more problematic that these models are used to study the linear stability of power-law films since a direct Orr-Sommerfeld analysis is not available. Indeed, as pointed out in \cite{RQ}, the divergence of the apparent viscosity prevented (especially in the case $0<n<1$), up to now (see thereafter), to write a suitable linearization of Navier-Stokes equations about a Nusselt type solution. A first order consistent shallow water model was derived by Amaouche et al \cite{ADB}, using the method of Ruyer Quil and Manneville. The first order models obtained in this way are not unique and moreover are not conservative, which can cause trouble in the presence of shocks. In the meantime, Fernandez-Nieto and the two authors \cite{FNV}  proposed an alternative model that is first order and conservative by a more direct approach based on the expansion of the fluid velocity and pressure field up to order one with respect to $\varepsilon$. Very recently, Ruyer Quil et al \cite{RQ} proposed a second order shallow water model in order to take into account of stream wise diffusion of momentum that plays an important role in the interaction of solitary waves when steady flow is unstable: here a regularization procedure was needed to apply Galerkin type method when $n<1$ and various inconsistency appeared in some terms found in the model.\\

\noindent
The purpose of this paper is two fold. On the one hand, we aim at deriving consistent shallow water models that are consistent up to order two. On the other hand, we want to validate rigorously this computation by comparing the spectral stability of constant states of the shallow water model and the spectral stability of the corresponding constant states (Nusselt flows) solution  of the Navier -Stokes equations. Note that up to now, there was no such direct study of the spectral 
stability of constant states for power law fluids down an incline since Orr Sommerfeld equations could not be derived. The main issue here, as pointed out in \cite{RQ}, is to take care of the divergence of the apparent viscosity. Indeed, if $n<1$, one cannot linearize $\tau=\mu_p\|D(\bf{u})\|^{n-1}D(\bf{u})$ about the Nusselt type solution when $D(\bf{u})=0$, i.e. at the free surface. Since the problem is only at the free surface and for the apparent viscosity, we have chosen to keep fluid strain rate as a (bounded or integrable) variable. Moreover, we introduce a (new) weaker form of the Cauchy momentum equations by integrating the Cauchy momentum equations with respect to the cross stream variable on an interval $(z, h(x,t))$ for all $(x,t)$. In that setting, we require less regularity for the solutions in the cross stream variable. Indeed, instead of assuming that solutions are bounded with bounded derivatives, we only need that solutions lie in $L^p_z((0, h), \R^m)$ functional spaces defined as follow
$$
\displaystyle
L^p_z((0, h);\R^m)=\{ v:\Omega_t\to \R^m\:|\:\forall (x,t)\:\int_0^{h(x,t)}|v(t,x,z)|^pdz<\infty\},
$$

\noindent
where $\Omega_t$ denotes the fluid domain at time $t$. As a consequence, our new formulation does not contain the free surface boundary conditions and we can formulate a generalization of Orr Sommerfeld equations to non Newtonian power law fluids $n\neq 1$. Moreover, we can propose with this approach a unified treatment of the shear thinning case $n<1$ and shear thickening case $n>1$. Though, depending on the case treated, we have chosen the classical constitutive law if $n>1$ and the inverse constitutive law if $n<1$. We have carried out a computation of the dispersion relation in the large wavelength regime up to order $3$ which is important to decide whether second order shallow water models are consistent: in particular, we recover and quantify the fact that the model derived in \cite{RQ} is not consistent except for $n=1$. 

In order to derive consistent second order shallow water models, one has to compute an  expansion of  the velocity field up to order two with respect to $\varepsilon$.  
This expansion is valid in a $L^p_z$ sense only under the assumption that $n>1/2$. This threshold for $n$ is important since it seems related to conditions for Navier-Stokes to be well posed \cite{MNR}: in particular, for some applications in ice sheets dynamic ($n=1/3$), one has to introduce a regularization of the power law and the linear stability and shallow water models that may result should depend drastically of the regularization process. Note that, in contrast to \cite{RQ}, in our approach, no distinction between the shear thinning case $(n<1)$ and the shear thickening case $(n>1)$ is needed. \\
  
\noindent
The paper is organized as follows: in section \ref{sec2}, we introduce the set of equations that describes the motion of a thin power-law film down an incline and write it in a non dimensional form and in the shallow water scaling.  In section \ref{sec3}, we move to the derivation of shallow water equations: we first compute an expansion of the fluid strain rate, pressure, velocity up to order $2$ and then deduce a second order shallow water model. Throughout the study, we have considered the natural no-slip condition at the bottom. Other boundary conditions can be considered as well; like the Navier slip condition which is used to study thin film flow down a porous inclined plane \cite{UMBR}. Finally, in section \ref{sec4}, we derive the Orr-Sommerfeld equations and compute an expansion of the dispersion law in the small wavenumber regime up to order $4$. We then compare with results obtained for various shallow water models. Finally, we conclude and address some perspectives to this work.

\section{\label{sec2} Governing equations}

\subsection{Power fluid constitutive law}

We find in the literature various parametrizations of power law fluids, which depend on the possible applications. Usually, one describes a power law fluid  (denoted PL in the sequel)  as a generalized Newtonian fluid. Then the constitutive law reads
\begin{equation}\label{pl_gn}
\displaystyle
\tau=\mu_p\|D({\bf u})\|^{n-1}D({\bf u}),
\end{equation}
where ${\bf u}=(u,w)^T$ is the fluid velocity, $D({\bf u})$ is the deformation tensor and $\tau$ is the fluid strain rate and $D({\bf u})$ is the deformation tensor which is defined as
$$
\displaystyle
D({\bf u})=\left(\begin{array}{cc} \displaystyle 2\partial_x u & \partial_z u+\partial_x w\\
                                                         \displaystyle \partial_z u+\partial_x w & 2\partial_z w
                                                         \end{array}\right),
$$
\noindent
whereas $\displaystyle\|D({\bf u})\|^2={\rm Tr}(D({\bf u})D({\bf u})^T)/2$. This law is known as the Ostwald-de Waele or Norton Hoff and was first formulated for applications in metallurgy. Note that in the shear thinning case $n<1$, the apparent viscosity $\mu_p\|D({\bf u})\|^{n-1}$ diverges and becomes unbounded at zero deformation rate. In order to deal with this singularity, an alternative approach consists in inverting the constitutive law so that the deformation tensor is a function of the fluid strain rate. Then, one has $\|\tau\|=\mu_p\|D({\bf u})\|^n$ and $\|D({\bf u})\|=(\mu_p)^{-1/n}\|\tau\|^{1/n}$. By inserting this relation into \eqref{pl_gn}, one finds
\begin{equation}\label{plgni}
\displaystyle
D({\bf u})=(\mu_p)^{-1/n}\|\tau\|^{1/n-1}\tau.
\end{equation}
\noindent
One recovers the Glen-Nye  formulation of the constitutive law for PL fluids by setting $A=(\mu_p)^{-1/n}$ and $n_*=1/n$:
\begin{equation}\label{gl}
\displaystyle
D({\bf u})=A\|\tau\|^{n_*-1}\tau.
\end{equation}
\noindent
Usually, the coefficient $n_*$ lies in $(2, 4)$ and for most glaciers $n_*=3$ which corresponds to $n=1/3$. We use the parameterization \eqref{pl_gn} of PL fluids to write the Cauchy momentum equations whereas we use (a rescaled version of) the parameterization \eqref{plgni} to carry out asymptotic expansions and to linearize Cauchy momentum equations.

\subsection{Navier-Stokes eqns for a thin PL film down an incline}

We consider an incompressible  power law fluid, with constant density $\rho$, flowing down an inclined plane, with a slope $\theta$, under the action of the gravity (here $g$ denotes the gravity acceleration).  The two dimensional Navier-Stokes equations read, in the fluid domain $\Omega(t)=\{(x,z)\,|\,0\leq z\leq h(x,t)\}$:
{\setlength\arraycolsep{1pt}
\begin{eqnarray}
\label{nsud}
\displaystyle
\rho\left(\partial_t u+u\partial_x u+w\partial_z u\right)&=&\rho g\sin(\theta)+\partial_x(\tau_{xx}-p)+\partial_z\tau_{xz},\\
\label{nswd}
\displaystyle
\rho\left(\partial_t w+u\partial_x w+w\partial_z w\right)&=&-\rho g\cos(\theta)+\partial_x\tau_{xz}+\partial_z(\tau_{zz}-p),\\
\label{nsincd}
\displaystyle
\partial_x u+\partial_z w&=&0.
\end{eqnarray}}
\noindent
This system is completed by the following boundary conditions at the free surface $z=h(x,t)$:
\setlength\arraycolsep{1pt}
\begin{eqnarray}
\label{bc1d}
\displaystyle
-\partial_x h(\tau_{xx}-p)|_{x=h(x,t)}+\tau_{xz}|_{z=h(x,t)}=-\sigma\frac{\partial_{xx}h\partial_x h}{(1+\partial_x h^2)^{3/2}},\\
\label{bc2d}
\displaystyle
-\partial_x h\,\tau_{xz}|_{z=h(x,t)}+(\tau_{zz}-p)|_{z=h(x,t)}=\sigma\frac{\partial_{xx}h}{(1+\partial_x h^2)^{3/2}},\\
\label{impd}
\displaystyle
\partial_t h+u|_{z=h(x,t)}\partial_x h=w|_{z=h(x,t)},
\end{eqnarray}
\noindent
whereas at the bottom $z=0$: $\displaystyle u|_{z=0}=w|_{z=0}=0$. Note that we take account of the surface tension: the capillary coefficient is denoted $\sigma$. The fluid strain rate $\tau$ is related  to the deformation rate either by \eqref{pl_gn} or \eqref{plgni}. Let us now write this system into a non dimensional form and in the shallow water scaling. For that purpose, we assume that the characteristic depth of the film is $H$, $U$ its characteristic velocity and $L$ is a characteristic longitudinal wavelength. The shallowness of the film is determined by the aspect ratio $\varepsilon$: in what follows, we will assume $\varepsilon\ll 1$. The variables $(x,z,t)$ and unknowns $(\tau, p, u,w, h)$ scale as:
$$
\begin{array}{ll}
\displaystyle
x=Lx',\quad z=Hz',\quad t=\frac{L}{U}t',\\
\displaystyle
u=Uu',\quad w=\varepsilon Uw',\quad p=\rho gHp',\quad \tau=\mu_p(\frac{U}{H})^n\tau',
\end{array}
$$
\noindent
whereas the deformation rate scales as $D({\bf u})=\frac{U}{H}D'({\bf u'})$ with
$$
\displaystyle
D'({\bf u})=\left(\begin{array}{cc} \displaystyle 2\varepsilon \partial_x u & \partial_z u+\varepsilon^2\partial_x w\\
                                                         \displaystyle \partial_z u+\varepsilon^2\partial_x w & 2\varepsilon\partial_z w
                                                         \end{array}\right),
$$
\noindent
We further introduce, respectively, the non dimensional Froude number, the Reynolds number and the Weber number:
$$
\displaystyle
F^2=\frac{U^2}{gH},\quad R_e=\frac{\rho H^nU^{2-n}}{\mu_p},\quad \quad W_e=\frac{\rho HU^2}{\sigma}.
$$
\noindent
Then, by dropping the $'$ and setting $\lambda=R_e\sin(\theta)/F^2$, one may write (\ref{nsud},\ref{nswd},\ref{nsincd}) with the boundary conditions (\ref{bc1d},\ref{bc2d},\ref{impd}) as
{\setlength\arraycolsep{1pt}
\begin{eqnarray}
\label{nsua}
\displaystyle
\partial_z\tau_{xz}+\lambda&=&\varepsilon R_e\left(\partial_t u+\partial_x u^2+\partial_z (uw)\right)+\varepsilon\partial_x\left(\frac{R_e p}{F^2}-\tau_{xx}\right),\\
\label{nswa}
\displaystyle
\partial_z(\tau_{zz}-\frac{R_e p}{F^2})-\lambda\cot(\theta)&=&-\varepsilon\partial_x\tau_{xz}+\varepsilon^2R_e\left(\partial_t w+\partial_x(uw)+\partial_z w^2\right),\nonumber\\
\displaystyle
\label{nsinca}
\partial_z w+\partial_x u&=&0.\nonumber
\end{eqnarray}}
\noindent
The power law \eqref{plgni} reads
\begin{equation}\label{plgnia}
\displaystyle
2\varepsilon\partial_x u=\left(\tau_{xx}^2+\tau_{xz}^2\right)^{\displaystyle\frac{1-n}{2n}}\tau_{xx},\quad
\partial_z u+\varepsilon^2\partial_x w=\left(\tau_{xx}^2+\tau_{xz}^2\right)^{\displaystyle\frac{1-n}{2n}}\tau_{xz}.
\end{equation}
\noindent
whereas $\tau_{zz}=-\tau_{xx}$.\\ 

Next, we set $\kappa=\varepsilon^2 F^2/W_e$: in order to see the influence of surface tension at first order with respect to $\varepsilon$, we assume that $\kappa=O(1)$. The boundary conditions (\ref{bc1d},\ref{bc2d},\ref{impd}) at the free surface are given by
{\setlength\arraycolsep{1pt}
\begin{eqnarray}
\label{bc1a}
\displaystyle
\tau_{xz}|_{z=h}-\varepsilon(\tau_{xx}-\frac{R_e p}{F^2})|_{z=h(x,t)}\partial_x h=-\kappa\frac{\varepsilon R_e\partial_x h\partial_{xx}h}{F^2(1+(\varepsilon\partial_x h)^2)^{3/2}},\\
\label{bc2a}
\displaystyle
(\tau_{zz}-\frac{R_e p}{F^2})|_{z=h(x,t)}-\varepsilon\tau_{xz}|_{z=h(x,t)}\partial_x h=\kappa\frac{ R_e\partial_{xx}h}{F^2(1+(\varepsilon\partial_x h)^2)^{3/2}},\\
\label{bc3a}
\displaystyle
\partial_t h+u|_{z=h(x,t)}\partial_x h=w|_{z=h(x,t)},
\end{eqnarray}}
\noindent
whereas the no-slip condition at the bottom is unchanged: $u|_{z=0}=w|_{z=0}=0$.\\

\noindent
Now we search for an Hilbert expansion of the fluid strain rate, velocity field with respect to the aspect ratio $\varepsilon\ll 1$: since the apparent viscosity diverges at the free surface, this expansion does not make sense in $L_z^{\infty}(0,h(x,t))$. Though this singularity is integrable and expansions will make sense in $L^p_z(0,h(x,t))$ spaces under some restrictions on $n\in(0, 1)$ and for suitable choices of $p\in[1,\,\infty[$. Hence we introduce a weaker formulation of (\ref{nsua}, \ref{nswa}, \ref{nsinca}) and  (\ref{bc1a}, \ref{bc2a}, \ref{bc3a}) that is adapted to this new functional spaces. This weak formulation is obtained by integrating (\ref{nsua}, \ref{nswa}) on the interval $(z,h(x,z))$ and by using the boundary conditions (\ref{bc1a}, \ref{bc2a}). One finds
{\setlength\arraycolsep{1pt}
\begin{eqnarray}
\displaystyle
\tau_{xz}&=&\lambda(h-z)-\frac{\kappa\varepsilon R_e}{F^2}\frac{\partial_x h\partial_{xx}h}{(1+(\varepsilon\partial_x h)^2)^{3/2}}+\varepsilon\partial_x\left(\int_z^h\tau_{xx}-\frac{R_e p}{F^2}d\tilde z\right)\nonumber\\
\label{nsuf}
\displaystyle
&&-\varepsilon R_e\left(\partial_t(\int_z^h u\,d\tilde z)+\partial_x(\int_z^hu^2\,d\tilde z)-uw\right),\\
\displaystyle
\tau_{xx}+\frac{R_e p}{F^2}&=&\lambda\cot(\theta)(h-z)-\frac{\kappa R_e}{F^2}\frac{\partial_{xx}h}{(1+(\varepsilon\partial_x h)^2)^{3/2}}-\varepsilon\partial_x(\int_z^h\tau_{xz}\,d\tilde z)\nonumber\\
\label{nswf}
\displaystyle
&&+\varepsilon^2R_e\left(\partial_t(\int_z ^h w\,d\tilde z)+\partial_x(\int_z^h uw\,d\tilde z)-w^2\right).
\end{eqnarray}}
\noindent
Any smooth solution of (\ref{nsua}, \ref{nswa}) and (\ref{bc1a}, \ref{bc2a}) is obviously a solution of (\ref{nsuf},\ref{nswf}). Moreover,  this weak formulation still holds for $(\tau_{xx},\tau_{xz},p)\in L^{1}_z(0, h(x,t))$ and $(u,w)\in L^2_z((0,h(x,t))$. Note that one can also write a weak form of \eqref{bc3a} by integrating the divergence free condition on $(z, h(x,t))$: one finds
\begin{equation}\label{cmassz}
\displaystyle
w=\partial_t h+\partial_x(\int_z^h u\,d\tilde z).
\end{equation}

\noindent
One purpose of this paper is to derive shallow water equations for $h$ and $q=\int_0^h u$ from the Navier-Stokes system in the shallow water scaling.  We find an exact system of evolution for $(h,q)$ by setting $z=0$ in \eqref{nsuf} and \eqref{cmassz} and by using the impermeability condition $w|_{z=0}=0$:
{\setlength\arraycolsep{1pt}
\begin{eqnarray}
\label{Cmass}
\displaystyle
\partial_t h+\partial_x(\int_0^h u\,dz)&=&0,\\
\label{Cmomentum}
\displaystyle 
\partial_t(\int_0^h u\,dz)+\partial_x\left(\int_0^h u^2+\frac{p}{F^2}\right)&=&\frac{1}{\varepsilon R_e}\left(\lambda h-\tau_{xz}|_{z=0}\right)+\partial_x(\int_0^h\frac{\tau_{xx}}{R_e}\,dz)\nonumber\\
\displaystyle
&&-\frac{\kappa}{F^2}\frac{\partial_x h\partial_{xx}h}{(1+(\varepsilon\partial_x h)^2)^{3/2}}.
\end{eqnarray}}
\noindent
The mass conservation law \eqref{Cmass} is written in a closed form with $(h,q)$ whereas one needs to write \eqref{Cmomentum} in a closed form. This latter step is done by
\begin{enumerate}
\item Computing an asymptotic expansion of $\tau_{xz},\tau_{xx},u,w,p$ with respect to $\varepsilon$,
\item Computing an expansion of $q=\int_0^h u\,dz$ with respect to $\varepsilon$,
\item Writing $\int_0^h u^2, \int_0^h p,\int_0^h \tau_{xx}$ and $\tau_{xz}|_{z=0}$ as function of $h,q$ up to a given order with respect to $\varepsilon$.
\end{enumerate}
\noindent
We will see that there are various choices  for this closure step but some of them are more natural and various constraints may also reduce the possibilities.

\section{\label{sec3} Derivation of shallow water models}

In this section, we  compute an expansion of the fluid strain rate, fluid pressure and the velocity field with respect to $\varepsilon$. Then we deduce lubrication models that are useful to study the stability of constant states of the full Cauchy momentum equations. Finally, we derive a consistent second order shallow water model for power law fluids. 

\subsection{Expansion of solutions to Cauchy momentum equations in the shallow water regime}

We assume that $(u,w)$, $(\tau,p)$ solution of (\ref{nsuf},\ref{nswf},\ref{nsinca},\ref{plgnia})   expand as
$$
\begin{array}{ll}
\displaystyle
(u,w)=(u^{(0)},w^{(0)})+\varepsilon(u^{(1)},w^{(1)})+\varepsilon^2(u^{(2)},w^{(2)})+O(\varepsilon^3),\\
\displaystyle
(\tau,p)=(\tau^{(0)},p^{(0)})+\varepsilon(\tau^{(1)},p^{(1)})+\varepsilon^2(\tau^{(2)},p^{(2)})+O(\varepsilon^3).
\end{array}
$$
\noindent
By collecting $O(\varepsilon^0)$ terms respectively in (\ref{nsuf},\ref{nswf}) and \eqref{plgnia}, one finds
$$
\begin{array}{ll}
\displaystyle
\tau^{(0)}_{xz}=\lambda(h(x,t)-z),\quad \tau^{(0)}_{xx}+\frac{R_e p^{(0)}}{F^2}=\lambda\cot(\theta)(h(x,t)-z)-\frac{\kappa R_e}{F^2}\partial_{xx}h,\\
\displaystyle
\tau^{(0)}_{xx}=0,\quad \partial_z u^{(0)}=\lambda^{1/n}(h(x,t)-z)^{1/n}.
\end{array}
$$
\noindent
As a consequence, the fluid pressure is hydrostatic:
$$
\displaystyle
 p^{(0)}=\cos(\theta)\left(h(x,t)-z\right)-\kappa\partial_{xx}h.
$$
\noindent
The fluid velocity $(u^{(0)},w^{(0)})$ vanishes at the bottom and is given by
{\setlength\arraycolsep{1pt}
\begin{eqnarray}
\displaystyle
u^{(0)}&=&\frac{n\lambda^{1/n}}{n+1}\left(h(x,t)^{1+1/n}-(h(x,t)-z)^{1+1/n}\right),\nonumber\\
\displaystyle
w^{(0)}&=&-\int_0^z\partial_x u^{(0)}d\tilde z\nonumber\\
\displaystyle
&=&-\lambda^{1/n}\left(\frac{h(x,t)^{1+1/n}}{n+1}+\frac{n}{n+1}(h(x,t)-z)^{1+1/n}-h(x,t)^{1/n}(h(x,t)-z)\right)\partial_x h.\nonumber
\end{eqnarray}}

\noindent
As quoted in \cite{RQ}, the apparent viscosity $(\partial_z u^{(0)})^{n-1}$ diverges at the free surface; though this singularity is integrable and asymptotic expansions hold in our mathematical framework.\\ 

\noindent
Note that one easily recovers the constant, Nusselt type, solutions of the Cauchy momentum equations by setting $h(x,t)=\bar h$:\begin{equation}\label{Nusselt}
\begin{array}{ll}
\displaystyle
\bar u=\frac{n\lambda^{1/n}}{n+1}\left(\bar h^{1+1/n}-(h\bar h-z)^{1+1/n}\right),\\
\displaystyle
\bar \tau_{xz}=\lambda(\bar h-z),\quad \frac{R_e}{F^2}\bar p=\lambda\cot(\theta)(\bar h-z),\\
\displaystyle
\bar\tau_{xx}=\bar\tau_{zz}=0,\quad \bar w=0.
\end{array}
\end{equation}
\noindent
Hence, by restricting the analysis to small amplitude motions $|h(x,t)-\bar h|\ll 1$, we expect that the asymptotic expansion shall provide a useful information on the stability of constant states under large scale perturbations: this is done in the last section of the paper.\\

\noindent
Next, we collect $O(\varepsilon)$ terms in (\ref{nsuf},\ref{nswf}) and (\ref{plgnia}):
{\setlength\arraycolsep{1pt}
\begin{eqnarray}
\displaystyle
\tau_{xz}^{(1)}&=&-R_e\left(\partial_t\int_z^h u^{(0)}d\tilde z+\partial_x\int_z^h (u^{(0)})^2d\tilde z-u^{(0)}w^{(0)}\right)\nonumber\\
\displaystyle
&&-\partial_x(\int_z^h\frac{R_e p^{(0)}}{F^2} d\tilde z)-\frac{\kappa R_e}{F^2}\partial_xh\partial_{xx}h,\\
\displaystyle
\frac{R_e p^{(1)}}{F^2}&=&-\partial_x(\int_z^h\tau_{xz}^{(0)})-\tau_{xx}^{(1)},\\
\displaystyle
\tau^{(1)}_{xx}&=&2(\tau_{xz}^{(0)})^{1-1/n}\partial_x u^{(0)}=2\lambda\left(h^{1/n}(h-z)^{1-1/n}-(h-z)\right)\partial_x h.
\end{eqnarray}}

\noindent
Hence, the correction $p^{(1)}$ to the fluid pressure is given by
$$
\displaystyle
\frac{R_e p^{(1)}}{F^2}=\lambda\left((h-z)-2h^{1/n}(h-z)^{1-1/n}\right)\partial_x h.
$$
\noindent
The weak formulation (\ref{nsuf},\ref{nswf}) holds only if $(\tau,p)\in L^1_z((0, h); M_2(\R)\times\R)$: we easily deduce from the expression of $\tau^{(1)}_{xx}$ that our calculations only make sense if $n>1/2$. In what follows, we will see that there is no additional restriction on $n$ in order to compute an expansion of the fluid strain rate and velocity field up to second order with respect to $\varepsilon$. Note that such a restriction on $n$ is also met when proving the well posedness of the Cauchy problem for power law fluids in bounded domains with periodic or Dirichlet (no-slip) boundary conditions \cite{MNR}.\\

\noindent
Now we can compute $\tau^{(1)}_{xz}$ (and thus $u^{(1)}$). Note that it contains time derivatives in $h$: it is standard in the derivation of thin film models that one can exchange time and spatial derivatives. More precisely, by using the mass conservation law \eqref{Cmass} which reads
$$
\displaystyle
\partial_t h+\partial_x(\int_0^h u^{(0)}dz)=O(\varepsilon).
$$
\noindent
and by using the expression for $u^{(0)}$, one finds
$$
\displaystyle
\partial_t h+\partial_x\left(\frac{n\lambda^{1/n}}{2n+1}h^{2+1/n}\right)=O(\varepsilon),
$$
\noindent
so that one can replace $\partial_t h$ by $-\lambda^{1/n}h^{1+1/n}\partial_x h$ in the expression for $\tau_{xz}^{(1)}$. A lengthly but straightforward computation that was implemented in a formal computation software provides the following expression for $\tau^{(1)}_{xz}$:
{\setlength\arraycolsep{1pt}
\begin{eqnarray}
\displaystyle
\tau^{(1)}_{xz}&=&R_e\lambda^{2/n}h^{1/n}\frac{(2n+1)h^{1+1/n}(h-z)-n(h-z)^{2+1/n}}{(2n+1)(n+1)}\partial_x h\nonumber\\
\displaystyle
&&+\kappa\frac{R_e}{F^2}(h-z)\partial_{xxx}h-\lambda\cot(\theta)(h-z)\partial_x h.\nonumber
\end{eqnarray}
}
\noindent
In particular, the tangential bottom strain rate expands as:
\begin{equation}\label{tau0}
\begin{array}{ll}
\displaystyle
\tau_{xz}|_{z=0}=\tau^{(0)}_{xz}|_{z=0}+\varepsilon\tau^{(1)}_{xz}|_{z=0}+O(\varepsilon^2),\\
\displaystyle
\tau_{xz}|_{z=0}=\lambda h+\varepsilon\left(\frac{\lambda^{2/n}h^{2+2/n}R_e}{2n+1}-\lambda h\cot(\theta)\right)\partial_x h+O(\varepsilon^2).
\end{array}
\end{equation}
\noindent
We complete the description of the velocity field at order one by computing successively $\partial_z u^{(1)}$, $u^{(1)}=\int_0^z\partial_z u^{(1)}d\tilde z$ and $w^{(1)}=-\int_0^z\partial_x u^{(1)}d\tilde z$. Indeed, by collecting $O(\varepsilon)$ term into \eqref{plgnia}, one finds
$$
\displaystyle
\partial_{z}u^{(1)}=\frac{1}{n}\left(\lambda(h-z)\right)^{1/n-1}\tau_{xz}^{(1)}
$$ 
\noindent
and $u^{(1)}, w^{(1)}$ easily follow, which completes the description of the velocity field up to order one. For the sake of conciseness of the presentation, we only provide the expression of interest to us, namely the correction to the discharge rate $q^{(1)}$:
{\setlength\arraycolsep{1pt}
\begin{eqnarray}
\displaystyle
q^{(1)}&=&\int_0^h(h-z)\partial_z u^{(1)}dz\nonumber\\
\displaystyle
&=&\left(\frac{2\lambda^{3/n-1}h^{3+3/n}R_e}{(2n+1)(3n+2)}-\frac{\lambda^{1/n}h^{2+1/n}\cot(\theta)}{2n+1}\right)\partial_x h+\frac{\kappa R_e}{(2n+1)F^2}\lambda^{1/n-1}h^{2+1/n}\partial_{xxx}h\nonumber,
\end{eqnarray}
}
\noindent
so that the full discharge rate $q$ expands as
{\setlength\arraycolsep{1pt}
\begin{eqnarray}
\displaystyle
q=\frac{nh^2(\lambda h)^{1/n}}{2n+1}&+&\varepsilon\left(\frac{2\lambda^{3/n-1}h^{3+3/n}R_e}{(2n+1)(3n+2)}-\frac{\lambda^{1/n}h^{2+1/n}\cot(\theta)}{2n+1}\right)\partial_x h\nonumber\\
\displaystyle
&+&\varepsilon\frac{\kappa R_e}{(2n+1)F^2}\lambda^{1/n-1}h^{2+1/n}\partial_{xxx}h+O(\varepsilon^2).\nonumber
\end{eqnarray}}
\noindent
Note that $\tau^{(1)}_{xx}$ does not play any role in the first order expansion of both $q^{(1)}$ and $\tau_{xz}^{(1)}|_{z=0}$ and thus there is no restriction on $n\in]0, 1[$ to write {\it first order} models. This is consistent with a previous paper by two of the authors \cite{FNV}: the expansion was carried out in $L^{\infty}_z(0, h)$ which resulted in a complicated treatment of expansion near the free surface where the apparent viscosity diverges. Here, our weaker  $L^1_z(0, h)$ formulation of the problem provides a simpler approach to compute a first order expansion of the solution and derive first order shallow water models.\\

\noindent
In contrast to \cite{FNV}, we are now in a position to derive second order expansion of both the tangential strain rate and discharge rate by using the new formulation (\ref{nsuf},\ref{nswf},\ref{plgnia}) of the Navier-Stokes equations under the restriction that $n>1/2$. Collecting $O(\varepsilon^2)$ terms in \eqref{plgnia}, one finds
$$
\displaystyle
\tau_{xx}^{(2)}=2\big(\lambda(h-z)\big)^{1-1/n}\partial_x u^{(1)}+\frac{n-1}{n}(\lambda(h-z))^{-1}\tau^{(1)}_{xz}\,\tau^{(1)}_{xx}.
$$
\noindent
Note that this function is $L^{1}_z(0, h)$ only if $n>1/2$ but  $\tau_{xx}^{(2)}$ does not play a role in the derivation of a second order shallow water model. Next, we compute $\tau_{xz}^{(2)}$ which, in turn, provides an expression for $u^{(2)}$ and $q^{(2)}$. One obtains $\tau^{(2)}_{xz}$ by collecting $O(\varepsilon^2)$ terms in \eqref{nsuf}: 
{\setlength\arraycolsep{1pt}
\begin{eqnarray}
\displaystyle
\tau_{xz}^{(2)}&=&-R_e\left(\partial_t\int_z^h u^{(1)}+2\partial_x\int_z^h u^{(0)}u^{(1)}-(u^{(0)}w^{(1)}+u^{(1)}w^{(0)})\right)\nonumber\\
\displaystyle
&&+\partial_x\left(\int_z^h\tau^{(1)}_{xx}-\frac{R_e p^{(1)}}{F^2}\right).
\end{eqnarray}}

\noindent
The first part of the expression of $\tau^{(2)}_{xz}$ which lies in front of $R_e$ is easily proved to be $L^{\infty}_z(0, h)$ whereas the last part is $O\left((h-z)^{1-1/n})\right)$ which is $L^{1}_z(0, h)$ provided that $n>1/2$. Next, by setting $z=0$ in the expression of $\tau^{(2)}_{xz}$, we obtain the second order correction to the bottom stress:

{\setlength\arraycolsep{1pt}
\begin{eqnarray}
\displaystyle
\tau^{(2)}_{xz}|_{z=0}&=&R_e^2\left(\frac{2\,{\lambda}^{\frac{4}{n}-1}\,{h}^{\frac{4}{n}+4}\,\left( 5\,{n}^{2}+14\,n+6\right) }{{\left( 2\,n+1\right) }^{2}\,\left( 3\,n+2\right) \,\left( 4\,n+3\right) }\right)\partial_{xx}h\nonumber\\
\displaystyle
&&-R_e\cot(\theta)\left(\frac{2\,{\lambda}^{\frac{2}{n}}\,{h}^{\frac{2}{n}+3}\,\left( n+2\right) }{\left( 2\,n+1\right) \,\left( 3\,n+2\right) }\right)\partial_{xx}h+\frac{\lambda\,{h}^{2}\,\left( 2\,n+3\right) }{2\,\left( 2\,n-1\right) } \partial_{xx}h\nonumber\\
\displaystyle
&&+R_e^2\left(\frac{2\,{\lambda}^{\frac{4}{n}-1}\,{h}^{\frac{4}{n}+3}\,\left( n+1\right) \,\left( 12\,{n}^{2}+46\,n+21\right) }{n\,{\left( 2\,n+1\right) }^{2}\,\left( 3\,n+2\right) \,\left( 4\,n+3\right) }\right)(\partial_x h)^2\nonumber\\
\displaystyle
&&-R_e\cot(\theta)\left(\frac{{\lambda}^{\frac{2}{n}}\,{h}^{\frac{2}{n}+2}\,\left( n+3\right) }{n\,\left( 2\,n+1\right) }\right)(\partial_x h)^2+\frac{\lambda\,h\,\left( 2\,n+3\right) }{2\,n-1}(\partial_x h)^2\nonumber\\
\displaystyle
&&+\frac{(n+1)\kappa\lambda^{2/n-1}h^{2/n}}{(2n+1)F^2}R_e^2\left(\frac{(1-n)(\partial_x h)^4}{n^3}+\frac{6h(\partial_x h)^2\partial_{xx}h}{n^2}+\frac{3h^2(\partial_{xx}h)^2}{n}\right)\nonumber\\
\displaystyle
&&+2\frac{\kappa\lambda^{2/n-1}h^{2/n}}{F^2}R_e^2\left(\frac{(2n+3)h^2\partial_x h\partial_{xxx}h}{n(2n+1)}+\frac{(n+2)h^3\partial_{xxxx}h}{(2n+1)(3n+2)}\right).\nonumber
\end{eqnarray}}

\noindent
Now, we compute successively $\partial_z u^{(2)}$, $u^{(2)}=\int_0^z \partial_z u^{(2)}d\tilde z$ and $q^{(2)}=\int_0^h(h-\tilde z)\partial_zu^{(2)}d\tilde z$. By collecting $O(\varepsilon^2)$ terms in \eqref{plgnia}, one finds
$$
\displaystyle
\partial_z u^{(2)}=-\partial_x w^{(0)}+\frac{1}{n}(\tau^{(0}_{xz})^{\displaystyle\frac{1}{n}-1}\tau^{(2)}_{xz}-\frac{n-1}{2n}(\tau^{(0)}_{xz})^{\displaystyle\frac{1}{n}-2}\left(\frac{(\tau^{(1)}_{xz})^2}{n}+(\tau^{(1)}_{xx})^2\right).
$$
\noindent
A careful inspection of the various terms in the expansion of $\partial_z u^{(2)}$ shows that $|\partial_z u^{(2)}|=O((h-z)^{-1/n})$ which is not $L^1_z((0, h))$. However, $u^{(2)}=\int_0^z \partial_z u^{(2)}=O((h-z)^{1-1/n})$ is integrable on $(0, h)$ and the second order correction to the discharge rate $q$ defined as $q^{(2)}=\int_0^h u^{(2)}dz$ is finite. A lengthly but straightforward computation yields the following expression for $q^{(2)}$:

{\setlength\arraycolsep{1pt}
\begin{eqnarray}
\displaystyle
q^{(2)}&=&R_e^2\left(\frac{4{\lambda}^{\frac{5}{n}-2}\,{h}^{\frac{5}{n}+5}\,\left(8n^2+25n+12 \right) }{{\left( 2\,n+1\right) }^{2}\,\left( 3\,n+2\right) \,\left( 4\,n+3\right) \,\left( 5\,n+4\right) }\right)\partial_{xx}h\nonumber\\
\displaystyle
&&-R_e\cot(\theta)\left( \frac{2{\lambda}^{\frac{3}{n}-1}\,{h}^{\frac{3}{n}+4}\,\left( 5n^2+14n+6\right) }{\left( 2\,n+1\right)^2 \,\left( 3\,n+2\right) \,\left( 4\,n+3\right) }     \right)\partial_{xx}h\nonumber\\
\displaystyle
&&+\frac{4{\lambda}^{\frac{1}{n}}\,{h}^{\frac{1}{n}+3}\,\left( n+2\right) }{3\,\left( 2\,n-1\right) \,\left( 3\,n+1\right) }\partial_{xx}h\nonumber\\
\displaystyle
&&+R_e^2\left(\frac{3\,{\lambda}^{\frac{5}{n}-2}\,{h}^{\frac{5}{n}+4}\,\left( n+1\right) \,\left( 35\,{n}^{2}+226\,n+120\right) }{2\,n\,{\left( 2\,n+1\right) }^{2}\,\left( 3\,n+2\right) \,\left( 4\,n+3\right) \,\left( 5\,n+4\right) }\right)(\partial_x h)^2\nonumber\\
\displaystyle
&&-R_e\cot(\theta)\left(\frac{8\,{\lambda}^{\frac{3}{n}-1}\,{h}^{\frac{3}{n}+3}}{n\,\left( 2\,n+1\right) \,\left( 3\,n+2\right) }  \right)(\partial_x h)^2\nonumber\\
\displaystyle
&&-\frac{{\lambda}^{\frac{1}{n}}\,{h}^{\frac{1}{n}+2}\,\left(10n^2+25n+7- 3(n-1)(2n-1){\cot}^2(\theta)\right) }{6\,n\,\left( 2\,n-1\right) \,\left( 2\,n+1\right) }(\partial_x h)^2.\nonumber\\
\displaystyle
&&+\frac{\kappa^2R_e^2}{F^4}\frac{(1-n)(\lambda h)^{1/n}h^2(\partial_{xxx}h)^2}{2n(2n+1)\lambda^2}+\frac{\kappa R_e(\lambda h)^{1/n}}{F^2}\frac{(n-1)h^2\partial_x h\partial_{xxx}h}{n(2n+1)\lambda}{\cot}(\theta)\nonumber\\
\displaystyle
&&+\frac{\kappa (n+1)R_e^2(\lambda h)^{3/n}}{(2n+1)\lambda^2F^2}\left(\frac{2(1-n)h(\partial_x h)^4}{(3n+2)n^3}+\frac{12h^2(\partial_x h)^2\partial_{xx}h}{(3n+2)n^2}+\frac{6h^3(\partial_{xx} h)^2}{(3n+2)n}\right)\nonumber\\
\displaystyle
&&+\frac{\kappa R_e^2(\lambda h)^{3/n}}{(2n+1)\lambda^2F^2}\left(\frac{2(3n+7)h^3\partial_x h\partial_{xxx}h}{n(3n+2)}+\frac{2(5n^2+14n+6)h^4\partial_{xxxx}h}{(2n+1)(3n+2)(4n+3)}\right)\nonumber
\end{eqnarray}}

\noindent
This completes the description of the fluid strain rate and velocity field up to order $2$ with respect to $\varepsilon$. We will use these expansions to provide various thin layer models both one equation (lubrication) models and two equations (shallow water) models.

\subsection{One equation-lubrication models}

In the previous section, we expanded the fluid strain rate and velocity field with respect to the aspect ratio $\varepsilon$ in the shallow water scaling $\varepsilon\to 0$, where the coefficients are functions of the fluid height $h$ and its spatial derivatives and the cross stream variable $z$. As a consequence, these quantities are completely determined if the fluid height $h$ is known. One derives an equation for the fluid height by considering the mass conservation law
$$
\displaystyle
\partial_t h+\partial_x q=0,\quad q=\int_0^h udz
$$
\noindent
and by substituting the expansion of $q$ with respect to $\varepsilon$. Indeed, by inserting an expansion of $q$ to order $j=0,1,2$:
\begin{equation}\label{lubn}
\displaystyle 
\partial_t h+\partial_x q^{(j)}=O(\varepsilon^{j+1}),\quad j=0,1,2,
\end{equation}
\noindent
and dropping $O(\varepsilon^{j+1})$ terms, one obtains a hierarchy of lubrication models consistent with the full Cauchy momentum equations and of order $j$. The first model of this hierarchy is a Burgers type equation
\begin{equation}\label{lub0}
\partial_t h+\partial_x\left(\frac{nh^2(\lambda h)^{1/n}}{2n+1}\right)=0,
\end{equation}
\noindent
which exhibits singularities in finite time. Though, it provides a useful information, namely the phase velocity of small perturbations of constant states $c(h)=(\lambda h)^{1/n}h$. The next model in the hierarchy, obtained for $j=1$ is a generalization of the Benney equation:
{\setlength\arraycolsep{1pt}
\begin{eqnarray}
\label{lub1}
\displaystyle
\partial_t h+\partial_x\left(\frac{nh^2(\lambda h)^{1/n}}{2n+1}\right)&=&\varepsilon\partial_x\left(\left(\frac{\lambda^{1/n}h^{2+1/n}{\rm cotan}(\theta)}{2n+1}-\frac{2\lambda^{3/n-1}h^{3+3/n}R_e}{(2n+1)(3n+2)}\right)\partial_x h\right)\nonumber\\
\label{lub1}
\displaystyle
&&-\kappa\varepsilon\frac{R_e\lambda^{1/n-1}}{(2n+1)F^2}\partial_x(h^{2+1/n}\partial_{xxx}h).
\end{eqnarray}
}

\noindent
Equation \eqref{lub1} was already derived formally in \cite{NM} and in \cite{FNV}. In the former paper, the authors proposed two equations shallow water models that are not consistent with \eqref{lub1} (even in the Newtonian case). As already noted in \cite{NM}, \cite{FNV}, the model \eqref{lub1} yields a stability criterion of constant states under small wavenumber perturbations. More precisely, by  linearizing \eqref{lub1} about the constant solution $h=1$, one finds
$$
\displaystyle
\partial_t \dot h+\lambda^{1/n}\partial_x \dot h=\frac{\varepsilon\lambda^{1/n}}{2n+1}\partial_x\left(({\rm cotan}(\theta)-\frac{2\lambda^{2/n-1}R_e}{3n+2})\partial_x \dot h\right)-\kappa\varepsilon\frac{R_e\lambda^{1/n-1}}{(2n+1)F^2}\partial_{xxxx}\dot h.
$$
\noindent
Then, the constant state $h=1$ is stable if the following criterion is satisfied
\begin{equation}\label{Re_c}
\displaystyle
R_e\leq \frac{3n+2}{2\lambda^{2/n-1}}{\rm cotan}(\theta)=Re_c.
\end{equation}
\noindent
If the mean velocity of the Poiseuille flow is chosen as the characteristic velocity, equation \eqref{Re_c} is a generalization of the classical stability criterion for Newtonian fluids:
$$
\displaystyle
R_e\leq \frac{5}{6}{\rm cotan}(\theta).
$$
\noindent
If \eqref{Re_c} is satisfied, the equation \eqref{lub1} is globally well posed for small amplitude solutions which decrease asymptotically to $0$. However, if this criterion is not satisfied, equation \eqref{lub1} is ill-posed and one has to consider a higher order model. The next model in the hierarchy is 
$$
\displaystyle
\partial_th+\partial_x\left(q^{(2)}(h,\partial_x h,\partial_{xx}h)\right)=0.
$$
However, the next correction term here is dispersive and would not stabilize \eqref{lub1}: one should then consider a third order model that would yield a Kuramoto-Sivashinsky type equation. However, this type of equations only makes sense near the instability threshold $0<Re-Re_c\ll 1$  which is limited for applications. 

In the last section of this paper, we relate the lubrication models \eqref{lubn} with the linear stability of constant states of the full Navier-Stokes equations by a direct analysis of the linearized Navier-Stokes equations, so called Orr-Sommerfeld equations (which are written here in a non classical form).

\subsection{Two equations-Shallow Water type models}

In this section, we propose an alternative to lubrication models and introduce two equations shallow water models with a source term that 
\begin{enumerate}
\item contain the lubrication models \eqref{lub0} and \ref{lub1} in the limit $\varepsilon\to 0$  if $Re<Re_c$,
\item are locally well posed if $Re>Re_c$ for order one models,
\item are globally well posed for second order models.
\end{enumerate}

The two equations-shallow water models are derived from the integral equations (\ref{Cmass},\ref{Cmomentum}). Equation \eqref{Cmass} is exact and already written in a closed form. There remains to write  \eqref{Cmomentum} in a closed form:
$$
\displaystyle
\partial_t q+\partial_x(\int_0^h u^2+\frac{p}{F^2}dz)=\frac{1}{\varepsilon R_e}(\lambda h-\tau_{xz}|_{z=0})+\partial_x(\int_0^h\tau_{xx}dz)-\frac{\kappa}{F^2}\frac{\partial_x h\partial_{xx}h}{(1+(\varepsilon\partial_x h)^2)^{3/2}}.
$$
\noindent
One obtains a order $j+1$ model by considering the truncated equation
\begin{equation}\label{cmom_n}
\displaystyle
\partial_t q+\partial_x(\int_0^h (\tilde u^{(j)})^2+\frac{\tilde p^{(j)}}{F^2})=\frac{1}{\varepsilon R_e}(\lambda h-\tilde\tau_{xz}^{(j+1)}|_{z=0})+\partial_x(\int_0^h \tilde\tau_{xx}^{(j)})-\frac{\kappa}{F^2}\frac{\partial_x h\partial_{xx}h}{(1+(\varepsilon\partial_x h)^2)^{3/2}},
\end{equation}
\noindent
with
\begin{equation}\label{clos1}
\displaystyle
(\tilde u^{(j)},\tilde\tau^{(j)},\tilde p^{(j)})=\sum_{k=0}^j \varepsilon^k(u^{(j)}, \tau^{(j)},p^{(j)}).
\end{equation}
\noindent
Then these functions are written as function of $(h,q)$ and their spatial derivatives: in this process, various choices are possible. For the inertial term, one writes
\begin{equation}\label{clos2}
\displaystyle
\int_0^h (\tilde u^{(j)})^2dz=\beta(n)\frac{q^2}{h}+I(h,\varepsilon,n)+O(\varepsilon^{j+1}),
\end{equation}
\noindent where the form of $I$ depends drastically on the choice of $\beta(n)$. The fluid pressure being hydrostatic, one has
\begin{equation}
\displaystyle
\int_0^h\tilde{p}^{j}(z)dz=P(h,n,\varepsilon)+O(\varepsilon^{j+1}).
\end{equation}
\noindent
Finally, we write the bottom stress as it is found in the literature e.g. in \cite{NM}:
\begin{equation}\label{clos3}
\displaystyle
\tilde\tau^{(j+1)}_{xz}|_{z=0}=\left(\frac{(2n+1)q}{nh^2}\right)^n+F_j(h,n,\varepsilon)+O(\varepsilon^{j+2}).
\end{equation}
\noindent
According to \cite{FNV}, one can write $F_0(h,n,\varepsilon)=\partial_x f(h,n)$ for the first order model ($j=0$). Moreover if we add the Galilean invariance as a constraint, one has to choose $\beta(n)=1$. This imposes the following form of the first order model:
{\setlength\arraycolsep{1pt}
\begin{eqnarray}
\displaystyle
\partial_t h+\partial_x q&=&0,\nonumber\\
\displaystyle
\partial_t q+\partial_x\Bigg(\frac{q^2}{h}+\frac{n^2\lambda^{2/n}h^{3+2/n}}{(3n+2)(2n+1)^2}&+&\frac{(3n+2)\lambda{\cot}(\theta)h^2}{4(2n+1)R_e}\Bigg)\nonumber\\
\displaystyle
&=&\frac{3n+2}{2\varepsilon(2n+1) R_e}\left(\lambda h-(q\frac{2n+1}{n\,h^2})^n\right)\nonumber\\
\displaystyle
&&+\frac{\kappa}{F^2}\frac{(3n+2)}{2(2n+1)}h\partial_{xxx}h.
\end{eqnarray}}

\noindent
This model is, up to $O(\varepsilon^2)$ terms, equivalent to the first order models found in \cite{ADB} and \cite{RQ} (in this latter paper, this is obtained by dropping streamwise diffusion terms):
{\setlength\arraycolsep{1pt}
\begin{eqnarray}
\displaystyle
\partial_t q+\frac{11n+6}{4n+3}\frac{q\partial_x q}{h}&-&3\frac{2n+1}{4n+3}\frac{q^2}{h^2}\partial_x h+\frac{(3n+2)}{2(2n+1)R_e}\lambda{\cot}(\theta)h\partial_x h\nonumber\\
\label{sw_nc}
\displaystyle
&=&\frac{3n+2}{2\varepsilon(2n+1) R_e}\left(\lambda h-(\frac{(2n+1)q}{n\,h^2})^n\right)\nonumber\\
\displaystyle
&&+\frac{\kappa}{F^2}\frac{(3n+2)}{2(2n+1)}h\partial_{xxx}h.
\end{eqnarray}}
\noindent
Note that this latter system is not written in a conservative form. Indeed by using the expansion of $q$:
$$
\displaystyle
q=\frac{n}{2n+1}(\lambda h)^{1/n}h^2+O(\varepsilon)=f(h)+O(\varepsilon)
$$
\noindent
in the convection and source terms of \eqref{sw_nc} and up to order $O(\varepsilon R_e)$, one can derive a general family of shallow water models in the form
$$
\begin{array}{ll}
\displaystyle
\partial_t h+\partial_x q=0,\\
\displaystyle
\partial_t q+a(h,q)\partial_x h+b(h,q)\partial_x q-k(h,q)\partial_{xxx}h=\frac{1}{\varepsilon R_e}\left(g(h)-\psi(h,q)\right).
\end{array}
$$
that are {\it all consistent up to order one} if the functions $a,b,k,g,\psi$ satisfy some constraints at $q=f(h)$. We will write these constraints explicitely in the derivation of shallow water models which are consistent up to order two.\\

Up to now, there was only one attempt to derive a second order shallow water model for thin power law film \cite{RQ}: however, on the one hand the model derived is not consistent with the case $n=1$ for fully nonlinear term whereas we shall prove in the next section that quasilinear ones are also not consistent except in the case $n=1$: indeed, as already mentioned in \cite{RQ}, this is due to the fact that correctors in the inertial terms are neglected.

We set $\beta=\varepsilon R_e$ and search the second order shallow water model in the form
{\setlength\arraycolsep{1pt}
\begin{eqnarray}
\displaystyle
\partial_t h+\partial_x q=0,\nonumber\\
\displaystyle
\partial_t q+a(h,q)\partial_x h+b(h,q)\partial_x q&=&\frac{1}{\beta}\left(g(h)-\psi(h,q)\right)+\kappa k(h,q)\partial_{xxx}h\nonumber\\
&+&\beta\left(P_1(h,q)\partial_{xx}h+P_2(h,q)(\partial_x h)^2+P_3(h,q)\partial_{xx}q\right)\nonumber\\
\displaystyle
&+&\kappa\beta\left(N_{3,2}(h,q)(\partial_{xxx}h)^2+N_{1,4}(h,q)(\partial_x h)^4\right)\nonumber\\
\displaystyle
&+&\kappa\beta\left(N_{12,21}(h,q)\partial_{xx}h(\partial_x h)^2+N_{2,2}(h,q)(\partial_{xx}h)^2\right)\nonumber\\
\displaystyle
&+&\kappa\beta\left(N_{4,1}(h,q)\partial_{xxxx}h+N_{31,11}(h,q)\partial_{xxx}h\partial_x h\right).\nonumber
\end{eqnarray}}
\noindent
In the limit $\beta\to 0$, this model provide an expansion of $q$ in the form
$$
\displaystyle
q=\tilde{q}^{(0)}+\beta q^{(1)}+\beta^{2}\tilde q^{(2)}+O(\beta^3)
$$
\noindent
The shallow water model is consistent with the Cauchy momentum equations up to order two with respect to $\varepsilon$ if the following constraints are satisfied:
$$
\displaystyle
q^{(0)}=\tilde{q}^{(0)},\quad q^{(1)}=R_e\tilde q^{(1)},\quad q^{(2)}=R_e^2\tilde q^{(2)}.
$$
\noindent
The first constraint $q^{(0)}=\tilde q^{(0)}$ implies $g(h)=\psi(h,f(h))$ with 
$$
\displaystyle
f(h)=q^{(0)}(h)=\frac{n}{2n+1}(\lambda h)^{1/n}h^2.
$$
\noindent
Next, the constraint $q^{(1)}=R_e\tilde q^{(1)}$  is satisfied only if: 
$$
\begin{array}{ll}
\displaystyle
\frac{f'(h)^2-a(h,f(h))-b(h,f(h))f'(h)}{\psi_q(h,f(h))}=2\frac{(\lambda h)^{3/n}h^3}{(2n+1)(3n+2)\lambda}-\frac{(\lambda h)^{1/n}h^2}{R_e(2n+1)}{\cot}(\theta),\\
\displaystyle
\frac{k(h,f(h))}{\psi_q(h,f(h))}=\frac{(\lambda h)^{1/n}h}{(2n+1)F^2\lambda}.
\end{array}
$$
\noindent
Finally, we consider the constraint $q^{(2)}=R_e^2\tilde q^{(2)}$ and $A(h)=\psi_q(h,f(h))$. First, we identify the term in front of $\partial_{xx}h$ in $q^{(2)}$ and $\tilde q^{(2)}$:
{\setlength\arraycolsep{1pt}
\begin{eqnarray}
\displaystyle
\frac{P_1+(\lambda h)^{1/n}hP_3}{A}&+&\frac{(2h(\lambda h)^{1/n}-b)((\lambda h)^{2/n}h^2-a-(\lambda h)^{1/n}hb)}{A^2}\nonumber\\
\displaystyle
&=&4\frac{(8n^2+25n+12)h^5(\lambda h)^{5/n}}{\lambda^2(5n+4)(4n+3)(3n+2)(2n+1)^2}\nonumber\\
\displaystyle
&&-2\frac{(5n^2+14n+6)(\lambda h)^{3/n}h^4}{R_e(4n+3)(3n+2)(2n+1)^2\lambda}{\cot}(\theta)\nonumber\\
\displaystyle
&&+\frac{4(n+2)(\lambda h)^{1/n}h^3}{3(3n+1)(2n-1)R_e^2}\nonumber
\end{eqnarray}}
\noindent
Next, by identifying the terms in front of $(\partial_x h)^2$, one obtains 
{\setlength\arraycolsep{1pt}
\begin{eqnarray}
\displaystyle
\frac{nP_2+(n+1)(\lambda h)^{1/n}P_3}{nA}&+&\frac{a+bh(\lambda h)^{1/n}-h^2(\lambda h)^{2/n}}{2A^3}\left(5h^2(\lambda h)^{2/n}-3hb(\lambda h)^{1/n}-a\right)\psi_{qq}\nonumber\\
\displaystyle
&+&\frac{a+bh(\lambda h)^{1/n}-h^2(\lambda h)^{2/n}}{2A^3}(4h(\lambda h)^{1/n}-2b)\psi_{hq}\nonumber\\
\displaystyle
&=&\frac{3(n+1)(35n^2+226n+120)h^4(\lambda h)^{5/n}}{2\lambda^2(5n+4)(4n+3)(3n+2)(2n+1)^2}\nonumber\\
\displaystyle
&&-8\frac{(\lambda h)^{3/n}h^3}{\lambda n(3n+2)(2n+1)R_e}{\cot}(\theta)\nonumber\\
\displaystyle
&&+\frac{(\lambda h)^{1/n}h^2(-6n^2+9n-3)}{6n(2n+1)(2n-1)R_e^2}{\cot}^2(\theta)\nonumber\\
\displaystyle
&&+\frac{(\lambda h)^{1/n}h^2(10n^2+13n+19)}{6n(2n+1)(2n-1)R_e^2}.\nonumber
\end{eqnarray}
}
\noindent
Then, by identifying the term in front of $\left(h_{xx}\right)^{2}$ in $q^{(2)},\tilde q^{(2)}$, one finds
$$
\displaystyle
\frac{N_{2,2}}{A}=\frac{1}{F^2}\left(6\frac{(n+1)(\lambda h)^{3/n}h^3}{n\lambda^2(3n+2)(2n+1)}-3\frac{(n+1)(\lambda h)^{2/n}h^2}{n(2n+1)\lambda A}\right).
$$
\noindent
By collecting the term in front of  $\left(h_{x}h_{xxx}\right)$ in $q^{(2)},\tilde q^{(2)}$, one obtains
{\setlength\arraycolsep{1pt}
\begin{eqnarray}
\displaystyle
\frac{N_{31,11}}{A}&=&2\frac{(3n+7)(\lambda h)^{3/n}h^3}{n\lambda^2(3n+2)(2n+1)F^2}+\frac{(n-1)(\lambda h)^{1/n}h^2}{n\lambda F^2(2n+1)R_e}{\cot}(\theta)\nonumber\\
\displaystyle
&+&\frac{(n+1-2n^2)a+(5n-2n^2)bh(\lambda h)^{1/n}+(2n^2+nh^3b_q-(9n+7)h^2)(\lambda h)^{2/n}}{n(2n+1)\lambda A F^2}.\nonumber
\end{eqnarray}
}

\noindent
Then, we collect the term in front of $\left(h_{x}\right)^{2}h_{xx}$ in $q^{(2)},\tilde q^{(2)}$:
$$
\displaystyle
\frac{N_{12,21}}{A}=\frac{1}{F^2}\left(12\frac{(n+1)(\lambda h)^{3/n}h^2}{n^2\lambda^2(3n+2)(2n+1)}-6\frac{(n+1)(\lambda h)^{2/n}h}{n^2\lambda (2n+1)AF^2}\right).
$$
\noindent
Next, we identify $(\partial_x h)^4$ in front of $q^{(2)},\tilde q^{(2)}$:
$$
\displaystyle
\frac{N_{1,4}}{A}=\frac{1}{F^2}\left(2\frac{(1-n^2)h(\lambda h)^{3/n}}{n^3\lambda^2(3n+2)(2n+1)}+\frac{(n^2-1)(\lambda h)^{2/n}}{n^3\lambda A}\right).
$$
\noindent
Finally, we identify respectively $\left(h_{xxx}\right)^{2}$ and $\partial_{xxxx}h$ terms in $q^{(2)},\tilde q^{(2)}$:
$$
\displaystyle
\frac{N_{3,2}}{A}=\frac{\kappa}{F^4}\left(\frac{(n-1)(\lambda h)^{1/n}h^2}{2n\lambda^2(2n+1)}-\frac{h^4(\lambda h)^{2/n}\psi_{qq}}{2(2n+1)^2A\lambda^2}\right),
$$
and
$$
\displaystyle
\frac{N_{4,1}}{A}=\frac{1}{F^2}\left(2\frac{(5n^2+14n+6)h^4(\lambda h)^{3/n}}{(4n+3)(3n+2)(2n+1)^2\lambda^2}+\frac{(b-2h(\lambda h)^{1/n})h^2(\lambda h)^{1/n}}{(2n+1)A\lambda}\right).
$$

\noindent
We search for a shallow water model that is consistent and is in the simpler and classical form: for that purpose, we try to cancel the fully nonlinear capillary terms $N_{i,j}$ which contain four derivatives. First, if one sets $N_{2,2}=0$, then $A=\psi_q(h,f(h))$ is given by
\[
A={\lambda}^{{\frac{n-1}{n}}}{h}^{-{\frac{1+n}{n}}}\frac{\left(3\, n+2\right)}{2}=\psi_{{\it q}}\left(h,\frac{n{h}^{2+\frac{1}{n}}{\lambda}^{\frac{1}{n}}}{2\, n+1}\right)
\]
\noindent
One possible choice for $\psi_q(h,q)$ is then
\[
\psi_{{\it q}}\left(h,q\right)={q}^{n-1}\left(2\, n+1\right)^{n}{n}^{-n+1}{h}^{-2\, n}\frac{\left(3\, n+2\right)}{2}
\]
\noindent
and by integrating with respect to $q$, one finds the natural form of the friction term for power law films

\[
\psi\left(h,q\right)=\frac{\left(3\, n+2\right)}{2}(\frac{(2n+1)q}{nh^2})^n.
\]
\noindent
The capillary coefficient $k$ is written as
\[
k(h,q)={\frac{\,\,\left(3\, n+2\right)h}{2F^2\,\left(2\, n+1\right)}}.
\]
\noindent
As a byproduct of our particular choice,  we also get $N_{{3,2}}=N_{{12,21}}=N_{{1,4}}=0$. In order to cancel $N_{{4,1}}$, one needs
\[
b(h,\frac{n{h}^{2+\frac{1}{n}}{\lambda}^{\frac{1}{n}}}{2\, n+1})=n{\lambda}^{\frac{1}{n}}{h}^{{\frac{1+n}{n}}}\frac{\left(11\, n+6\right)}{8\,{n}^{2}+10\, n+3}
\]
and consequently, a suitable choice is
\[
b(h,q)=\frac{q}{h}\frac{\left(11\, n+6\right)}{4\, n+3}.
\]
\noindent
The consistency of the shallow water model at first order yields:
\[
\displaystyle
a(h,\frac{n{h}^{2+\frac{1}{n}}{\lambda}^{\frac{1}{n}}}{2\, n+1})=-\frac{3\,{\lambda}^{\frac{2}{n}}{h}^{2\,{\frac{1+n}{n}}}{n}^{2}}{8\,{n}^{2}+10\, n+3}+\frac{12{n}^{2}+17\, n+6\,}{2\left(8\,{n}^{2}+10\, n+3\right)}\frac{\lambda h}{R_e}{\cot}(\theta).
\]
\noindent
One possible choice is 
$$
\displaystyle
a(h,q)=-\frac{3(2n+1)}{4n+3}\frac{q^2}{h^2}+\frac{3n+2}{2(2n+1)}\frac{\lambda h}{R_e}{\cot}(\theta).
$$
\noindent
We also get $N_{{31,11}}=0$. Finally, there remains to determine $P_i$: one suitable choice is
{\setlength\arraycolsep{1pt}
\begin{eqnarray}
\displaystyle
P_2(h)&=&-\frac{9n(n+1)(5n-4)\lambda^{-1+4/n}h^{3+4/n}}{4(2n+1)^2(4n+3)(5n+4)(3n+2)},\nonumber\\
\displaystyle
P_3(h)&=&\frac{1}{R_e^2}\frac{(10n^2+25n+7)(3n+2)}{12(2n-1)(2n+1)(n+1)}(\frac{(2n+1)q}{nh^2})^{n-1},\nonumber\\
\displaystyle
P_1(h)&=&\frac{2n^2(1-n)\lambda^{-1+4/n}h^{4+4/n}}{(2n+1)^2(4n+3)(5n+4)(3n+2)}\nonumber\\
\displaystyle
&&+\frac{1}{R_e^2}\frac{n(3n+2)(9+10n-29n^2-14n^3)q}{12(2n-1)(2n+1)^2(3n+1)(n+1)h}(\frac{(2n+1)q}{nh^2})^{n-1}
\end{eqnarray}}

\noindent
Note that the first order consistent model found in \cite{ADB} and \cite{RQ} is precisely the one for which the correction terms due to surface tension vanish. The second order shallow water model reads
{\setlength\arraycolsep{1pt}
\begin{eqnarray}
\displaystyle
\partial_t q+\frac{11n+6}{4n+3}\frac{q\partial_x q}{h}&-&3\frac{2n+1}{4n+3}\frac{q^2}{h^2}\partial_x h+\frac{(3n+2)}{2(2n+1)R_e}\lambda{\cot}(\theta)h\partial_x h\nonumber\\
\label{sw_ncv}
\displaystyle
&=&\frac{3n+2}{2(2n+1)\beta}\left(\lambda h-(\frac{(2n+1)q}{n\,h^2})^n\right)+\frac{\kappa}{F^2}\frac{(3n+2)}{2(2n+1)}h\partial_{xxx}h\nonumber\\
\displaystyle
&&+\beta\left(P_1(h)\partial_{xx}h+P_2(h)(\partial_x h)^2+P_3(h)\partial_{xx}q\right).\nonumber
\end{eqnarray}}

\noindent
Note that $P_1$ and $P_3$, which play a role in the linear stability of constant states has $2n-1$ in the denominator and only make sense if $n>1/2$. Once this condition satisfied there is no distinction between the shear thinning case $n<1$ and shear thickening case $n>1$.

\section{\label{sec4} Direct stability analysis}

In this section, we study rigorously the linear stability of constant Nusselt flows without any regularization of the apparent viscosity. We show that the expansion of velocity field and fluid strain rate carried out previously in the shallow water/small wavenumber regime provide naturally an expansion of eigenfunctions and eigenvalues of the Orr-Sommerfeld equations in the small wave number regime. In particular, we show that the dispersion relation associated to linearized lubrication equations about the constant state $h=1$ yields the expansion of the eigenvalue $\lambda\approx 0$ of the full problem with respect to the wavenumber $k\to 0$. This analysis is valid up to order $2$, the order of the discharge rate expansion carried out previously: for completeness, we have computed an expansion of the eigenvalue $\lambda\approx 0$ up to order four with respect to $k$ in order to generalize the classical results for newtonian fluids.

\subsection{Rigorous formulation of the Orr-Sommerfeld equations}

In this section, we write rigorously the linearized Cauchy momentum equations about a Nusselt solution: in this case, the apparent viscosity diverges at the free surface. Though, by using our weak formulation and keeping the strain rate as an unknown, we remove this singularity. We start from the non dimensional Cauchy momentum equations (\ref{nsua},\ref{plgnia}) completed by the boundary conditions (\ref{bc1a},\ref{bc2a},\ref{bc3a}) and fix $\varepsilon=1$\footnote{At this stage, we do not focus on a particular wavenumber regime}. For the sake of conciseness, we have neglected surface tension in this discussion.\\ 

\noindent
In what follows, we study the stability of the constant (in $x$ and $t$) Nusselt solution
$$
\begin{array}{ll}
\displaystyle
h_0=1,\quad \tau_{0,xz}=\lambda(1-z),\quad \tau_{0,xx}=\tau_{0,zz}=0,\quad  p_0=c(1-z),\\
\displaystyle
W_0=0,\quad U_0=\frac{n\lambda^{1/n}}{n+1}\left(1-(1-z)^{1+1/n}\right).
\end{array}
$$
\noindent
We linearize the Cauchy momentum equations about this constant solution
{\setlength\arraycolsep{1pt}
\begin{eqnarray}
\label{nsul}
\displaystyle
\partial_t u+U_0\partial_x u+U_0'w+\frac{\partial_x p}{F^2}&=&\frac{1}{R_e}\left(\partial_x\tau_{xx}+\partial_z\tau_{xz}\right),\\
\label{nswl}
\displaystyle
\partial_t w+U_0\partial_x w+\frac{\partial_z p}{F^2}&=&\frac{1}{R_e}\left(\partial_x\tau_{xz}+\partial_z\tau_{zz}\right),\\
\label{nsincl}
\displaystyle
\partial_x u+\partial_z w&=&0.
\end{eqnarray}}
\noindent
The linearized boundary conditions at the free surface are written
\begin{equation}\label{bc_lin}
\begin{array}{ll}
\displaystyle
\tau_{xz}|_{z=1}=\lambda\tilde{h},\quad \frac{p}{F^2}|_{z=1}-\frac{\tau_{zz}}{R_e}|_{z=1}=\frac{c\tilde{h}}{F^2},\\
\displaystyle
\partial_t \tilde{h}+\frac{n\lambda^{1/n}}{n+1}\partial_x\tilde{h}=w|_{z=1}.
\end{array}
\end{equation}
\noindent 
The constant solution has a singular apparent viscosity: in order to write the linearized problem, we invert the relation $\tau=\|D(u)\|^{n-1}D(u)$ as $D(u)=\|\tau\|^{1/n-1}\tau$. Then, we linearize this constitutive law:
\begin{equation}\label{pl_lin}
\displaystyle
\left(\lambda(1-z)\right)^{\displaystyle\frac{1-n}{n}}\tau_{xx}=2\partial_x u,\quad \frac{1}{n}\left(\lambda(1-z)\right)^{\displaystyle\frac{1-n}{n}}\tau_{xz}=\partial_z u+\partial_x w.
\end{equation}

\noindent
Since the velocity field is divergence free, there is a potential function $\phi$ so that
$$
\displaystyle
u=-\partial_z\phi,\quad w=\partial_x \phi.
$$ 
\noindent
The system (\ref{nsul},\ref{nswl},\ref{nsincl}) and (\ref{pl_lin}) with the boundary conditions (\ref{bc_lin}) are homogeneous in time and space: we then search for solutions of this linearized system in the form 
$$
\displaystyle
(\phi,\sigma,p):=e^{ik\Lambda t+ikx}(\phi,\sigma,p).
$$ 
\noindent
Then, one finds
\begin{equation}\label{nscl}
\begin{array}{ll}
\displaystyle
\tau_{xz}'=ik\left(\frac{R_e p}{F^2}-\tau_{xx}+R_e(U_0'\phi-(\Lambda+U_0)\phi')\right),\nonumber\\
\displaystyle
\frac{R_e p'}{F^2}-\tau_{zz}'=ik\left(\tau_{xz}-ik\,R_e(\Lambda+U_0)\phi\right),
\end{array}
\end{equation}
\noindent
whereas the boundary conditions read
\begin{equation}\label{bccl}
\begin{array}{ll}
\displaystyle
\tau_{xz}|_{z=1}=\lambda\tilde{h},\quad \frac{R_e p}{F^2}|_{z=1}-\tau_{zz}|_{z=1}=\lambda{\cot}(\theta)\tilde{h},\\
\displaystyle
\phi|_{z=1}=(\Lambda+\frac{n\lambda^{1/n}}{n+1})\tilde{h},\quad \phi|_{z=0}=\phi'|_{z=0}=0.
\end{array}
\end{equation}
\noindent
The linearized constitutive law reads
\begin{equation}\label{plcl}
\displaystyle
\left(\lambda(1-z)\right)^{\displaystyle\frac{1-n}{n}}\tau_{xx}=-2ik\phi',\quad -\frac{1}{n}\left(\lambda(1-z)\right)^{\displaystyle\frac{1-n}{n}}\tau_{xz}=\phi''+k^2\phi.
\end{equation}
\noindent
Following  the strategy of the previous section, we integrate \eqref{nscl} over the interval $(z, 1)$ and use the boundary conditions \eqref{bccl}: this yields
{\setlength\arraycolsep{1pt}
\begin{eqnarray}
\label{nsulf}
\displaystyle
\tau_{xz}&=&\lambda\tilde h-ik\int_z^1\left(\frac{R_e p}{F^2}-\tau_{xx}+R_e(U'_0\phi-(\Lambda+U_0)\phi')\right)d\tilde z,\\
\label{nswlf}
\displaystyle
\frac{R_e p}{F^2}+\tau_{xx}&=&\lambda{\cot}(\theta)\tilde h-ik\int_z^1\tau_{xz}d\tilde z-k^2\int_z^1\,R_e(\Lambda+U_0)\phi d\tilde z.
\end{eqnarray}}
\noindent
This comes with
{\setlength\arraycolsep{1pt}
\begin{eqnarray}
\label{txl}
\displaystyle
\tau_{xx}&=&-2ik(\lambda(1-z))^{1-1/n}\phi',\\
\label{phil}
\displaystyle
\phi&=&-\int_0^z\int_0^{\tilde z}\frac{1}{n}(\lambda(1-y))^{1/n-1}\tau_{xz}+k^2\phi(y)dyd\tilde z.
\end{eqnarray}}
\noindent
In particular, one has
$$
\displaystyle
\phi|_{z=1}=-\int_0^1\frac{\lambda^{1/n-1}}{n}(1-\tilde{z})^{1/n}\tau_{xz}+k^2(1-\tilde z)\phi(\tilde z)d\tilde z.
$$

\subsection{Connection with the shallow water expansion}

In what follows, we will consider the low wavenumber regime $k\to 0$ which corresponds to the shallow water regime for the Navier Stokes system: we shall prove that Orr Sommerfeld equations are a linearized version of the ``weak'' formulation of Navier-Stokes equations and that expansion of eigenvalues and eigenfunctions of Orr Sommerfeld are obtained directly by the linearized iterative scheme.

In the shallow water scaling, $\tau,u,w,p$ are functions of $h$ and its partial derivatives and one can compute an expansion with respect to $\varepsilon$ up to order $2$:
$$
\displaystyle
(\tau,u,w,p)(z,x,t)=\sum_{k=0}^2\varepsilon^k\,(\tau^{(k)},u^{(k)},w^{(k)},p^{(k)})(z,h(x,t))+O(\varepsilon^3).
$$
\noindent
We denote 
$$
\begin{array}{ll}
\displaystyle
(T_0,U_0,P_0)=(\tau^{(0)}_{xz}, u^{(0)}, p^{(0)})(z,1),\\
\displaystyle
T_0=\lambda(1-z),\quad U_0=n\lambda^{1/n}(1-(1-z)^{1+1/n})/(n+1),\quad \frac{R_e\,P_0}{F^2}=\lambda{\cot}(\theta)(1-z).
\end{array}
$$

\noindent
By differentiating the weak formulation of Cauchy momentum equations (\ref{nsuf},\ref{nswf},\ref{cmassz}) and \eqref{plgnia} about $h=1$, one proves that 
$$
(\tilde\tau,\tilde u,\tilde w,\tilde p)=d(\tau,u,w,p)|_{h=1}(\tilde h),
$$
 \noindent
 satisfies the problem: 
 $$
 \displaystyle
 2\varepsilon\partial_x \tilde u=T_0^{\frac{1-n}{n}}\tilde \tau_{xx},\quad \partial_z \tilde u+\varepsilon^2\partial_x\tilde w=\frac{1}{n}T_0^{\frac{1-n}{n}}\tilde\tau_{xz},\quad \tilde u|_{z=0}=0.
 $$
\noindent
and
{\setlength\arraycolsep{1pt}
\begin{eqnarray}
\displaystyle
\tilde\tau_{xz}&=&\lambda\tilde h+\varepsilon\partial_x\left(\int_z^{1}\tilde \tau_{xx}-\frac{R_e}{F^2}(\tilde p+P_0(1)\tilde h\right)\nonumber\\
\displaystyle
&&-\varepsilon R_e\left(\partial_t(\int_z^1\tilde u+U_0(1)\tilde h)+\partial_x(U_0^2(1)\tilde h+2\int_z^1 U_0\tilde u)-U_0\tilde w\right),\nonumber\\
\displaystyle
\tilde\tau_{xx}+\frac{R_e\tilde p}{F^2}&=&\lambda{\cot}(\theta)\tilde h-\varepsilon\partial_x\left(\int_z^1\tilde{\tau}_{xz}+T_0(1)\tilde h\right)\nonumber\\
\displaystyle
&&+\varepsilon^2 R_e\left(\partial_t(\int_z^1\tilde w)+\partial_x(\int_z^1U_0\tilde w)\right),\nonumber
\end{eqnarray}}
\noindent
whereas the linearized mass conservation law \eqref{cmassz} is written as
$$
\displaystyle
\tilde w=\partial_t\tilde h+\partial_x\left(\int_z^1\tilde u+U_0(1)\tilde h\right),\quad \tilde{w}|_{z=0}=0.
$$
\noindent
Then, it is a straightforward computation to prove that
{\setlength\arraycolsep{1pt}
\begin{eqnarray}
\displaystyle
\tilde{\tau}_{xz}&=&\lambda\tilde h+\varepsilon\partial_x\left(\int_z^1\tilde\tau_{xx}-\frac{R_e\tilde p}{F^2}\right)\nonumber\\
\displaystyle
&&-\varepsilon R_e\left(\partial_t(\int_z^1\tilde u)+\partial_x(\int_z^1 U_0\tilde u+U'_0\tilde w)\right),\nonumber\\
\displaystyle
\tilde{\tau}_{xx}+\frac{R_e\tilde p}{F^2}&=&\lambda{\cot}(\theta)\tilde h-\varepsilon\partial_x(\int_z^1\tilde{\tau}_{xz})\nonumber\\
\displaystyle
&&+\varepsilon R_e\left(\partial_t(\int_z^1\tilde w)+\partial_x(\int_z^1U_0\tilde w)\right)\nonumber
\end{eqnarray}} 
\noindent
These equations are homogeneous in $x$ and $t$: we search for solutions in the form of plane waves 
$$
\displaystyle
(\tilde{\tau},\tilde u,\tilde w,\tilde p):=e^{i\xi \Lambda t+i\xi x}(\tau(z),-\phi'(z),i\xi\phi(z),p),
$$
\noindent
which yields precisely the linearized Orr-Sommerfeld equations \eqref{nsulf},\eqref{nswlf},\eqref{plcl} and \eqref{bccl} by setting $k=\varepsilon\xi$. As a result, one obtains directly an expansion with respect to $k$ of eigenvalues $(\Lambda,\phi,\tau)$  simply by identifying the coefficient in the Taylor expansion with respect to $\varepsilon$
{\setlength\arraycolsep{1pt}
\begin{eqnarray}
\displaystyle
\Lambda&=&\Lambda^{(0)}+k\Lambda^{(1)}+k^{2}\Lambda^{(2)}+O(k^3)\\
\displaystyle
&=&\Lambda^{(0)}+\varepsilon\xi\Lambda^{(1)}+\varepsilon^2\xi^2\Lambda^{(2)}+O(\varepsilon^{3}).
\end{eqnarray}}
\noindent
Indeed, the expansion of $\Lambda$ is easily obtained through the mass conservation law
$$
\displaystyle
\partial_t \tilde h+\partial_x(\int_z^1\tilde u+U_0(1)\tilde h)=0\quad\Longrightarrow \Lambda\tilde h+\int_0^1\tilde u+U_0(1)\tilde h=0.
$$
\noindent
Recall that $q=\int_0^h u$ so that $\displaystyle \tilde q=dq|_{h=1}\tilde h=\int_0^1\tilde u+U_0(1)\tilde h$. Moreover, we have proved in the previous section that $q$ expands as
$$
\displaystyle
q=q^{(0)}(h)+\varepsilon q^{(1)}(h)\partial_x h+\varepsilon^2 (q^{(2)}(h)\partial_{xx}h+r^{(2)}(h)(\partial_x h)^2)+O(\varepsilon^3),
$$
\noindent
so that $\tilde q$ is given by
$$
\displaystyle
\tilde q=(q^{(0)})'(1)\tilde h+\varepsilon q^{(1)}(1)\partial_x\tilde h+\varepsilon^2 q^{(2)}(1)\partial_{xx}\tilde h+O(\varepsilon^2)
$$
\noindent
From this expansion and by using the relation $\Lambda+\tilde q=0$, one finds
$$
\displaystyle
\Lambda^{(0)}=(q^{(0)})'(1),\quad \Lambda^{(1)}=-iq^{(1)},\quad \Lambda^{(2)}=q^{(2)}
$$ 

\noindent
This yields
$$
\displaystyle
\Lambda^{(0)}=-\lambda^{1/n},\quad \Lambda^{(1)}=-i\left(\frac{2\lambda^{3/n-1}R_e}{(2n+1)(3n+2)}-\frac{ \lambda^{1/n}{\cot}(\theta) }{2n+1}\right),
$$
and
{\setlength\arraycolsep{1pt}
\begin{eqnarray}
\displaystyle
\Lambda^{(2)}&=&R_e^2\left(\frac{4\,{\lambda}^{\frac{5}{n}-2}\,\left( 8\,{n}^{2}+25\,n+12\right) }{{\left( 2\,n+1\right) }^{2}\,\left( 3\,n+2\right) \,\left( 4\,n+3\right) \,\left( 5\,n+4\right) }\right)\nonumber\\
\displaystyle
&&-R_e{\cot}(\theta)\left(\frac{2\,{\lambda}^{\frac{3}{n}-1}\,\left( 5\,{n}^{2}+14\,n+6\right) }{{\left( 2\,n+1\right) }^{2}\,\left( 3\,n+2\right) \,\left( 4\,n+3\right) }
\right)\nonumber\\
\displaystyle
&&+\frac{4\,{\lambda}^{\frac{1}{n}}\,\left( n+2\right) }{3\,\left( 2\,n-1\right) \,\left( 3\,n+1\right) }
\end{eqnarray}}

\noindent
Similarly, one can obtain an expansion of the potential $\phi$ and fluid strain $\tau$ up to order $2$ with respect to $\varepsilon$. 

\subsection{Higher order correction to Orr Sommerfeld\\ eigenvalues/eigenfunctions}

We stopped the full non linear iterative scheme up to order 2 both from applications point of view (it is sufficient to derive second order shalllow water mdels) and due to some limitations due to the integrability of nonlinear terms in the expansion. However, at the linear level, one can continue the expansions one step further. As an application, one could deduce a consistent amplitude equation of Kuramoto-Sivashinsky type near the instability threshold.

Recall that we expand $\tau,\phi,p$ with respect to $k(=\varepsilon\xi)$ as $k\to 0$:
$$
\displaystyle
(\tau,\phi,p)=(\tau^{(0)},\phi^{(0)},p^{(0)})+k(\tau^{(1)},\phi^{(1)},p^{(1)})+k^2(\tau^{(2)},\phi^{(2)},p^{(2)})+O(k^3).
$$
\noindent
Collecting $O(k^0)$ terms into (\ref{nsulf},\ref{nswlf}) and (\ref{txl}) yields
$$
\displaystyle
\tau^{(0)}_{xz}=\lambda\tilde h,\quad \tau^{(0)}_{xx}=0,\quad \frac{R_e p^{(0)}}{F^2}=\lambda{\cot}(\theta)\tilde h.
$$
\noindent
We then deduce that
$$
\displaystyle
\phi^{(0)}=\lambda^{1/n}\left(-\frac{1}{n+1}+(1-z)-\frac{n}{n+1}(1-z)^{1+1/n}\right).
$$
\noindent
This provides 
$$
\displaystyle
\phi^{(0)}|_{z=1}=\Lambda^{(0)}+\frac{n\lambda^{1/n}}{n+1}=\frac{n\lambda^{1/n}}{n+1}-\lambda^{1/n},
$$
\noindent
so that $\Lambda^{(0)}=-\lambda^{1/n}$. Next, we compute higher order corrections. For that purpose, we  collect $O(k)$ terms into \eqref{nsulf} and \eqref{nswlf}: this yields
{\setlength\arraycolsep{1pt}
\begin{eqnarray}
\displaystyle
\tau_{xz}^{(1)}&=&-i\int_z^1\left(\frac{R_e p^{(0)}}{F^2}+R_e(U_0'\phi^{(0)}-(\Lambda^{(0)}+U_0)(\phi^{(0)})'\right)d\tilde z\nonumber\\
\displaystyle
&=&-i\left(\lambda{\rm cotan}(\theta)(1-z)+\frac{R_en\lambda^{2/n}}{n+1}(\frac{(1-z)^{2+1/n}}{2n+1}-\frac{1-z}{n})\right).\nonumber
\end{eqnarray}}
\noindent
Then, one obtains
$$
\displaystyle
(\phi^{(1)})''=i\lambda^{1/n}\left(\frac{{\cot}(\theta)}{n}(1-z)^{1/n}+\frac{R_e\lambda^{2/n-1}}{n+1}(\frac{(1-z)^{1+2/n}}{2n+1}-\frac{(1-z)^{1/n}}{n})\right).
$$
\noindent
By integrating twice this equation with the boundary conditions $\phi|_{z=0}=\phi'|_{z=0}=0$, one finds
{\setlength\arraycolsep{1pt}
\begin{eqnarray}
\displaystyle
\phi^{(1)}&=&i\lambda^{1/n}\left(\frac{{\cot}(\theta)}{n+1}(z-\frac{n(1-(1-z)^{2+1/n})}{2n+1})\right)\nonumber\\
\displaystyle
&&+iR_e\frac{\lambda^{3/n-1}}{(n+1)^2}\left(\frac{n(1-(1-z)^{2+1/n})}{2n+1}-z+\frac{n}{4n+2}(z-\frac{n(1-(1-z)^{3+2/n})}{3n+2})\right).\nonumber
\end{eqnarray}}
\noindent
This yields 
$$
\displaystyle
\Lambda^{(1)}=i\frac{\lambda^{1/n}}{2n+1}\left({\cot}(\theta)-\frac{2R_e\lambda^{2/n-1}}{3n+2}\right).
$$
\noindent
As a result, we obtain the following expansion
$$
\displaystyle
\Lambda(k)=-ik\lambda^{1/n}-k^2\frac{\lambda^{1/n}}{2n+1}\left({\cot}(\theta)-\frac{2R_e\lambda^{2/n-1}}{3n+2}\right)+O(k^3).
$$
\noindent
which provides the stability criterion under small wavenumber perturbations
$$
\displaystyle
R_e\leq \frac{(3n+2)\lambda^{1-2/n}}{2}{\cot}(\theta).
$$
\noindent
This criterion was already derived in \cite{NM} and \cite{FNV}  from a consistent lubrication theory. However, this is the first rigorous result that is derived directly from the Cauchy momentum equations through a Orr-Sommerfeld type analysis. The key point was to introduce the weak formulation so that one can obtain a well posed linearized problem. This approach is also suitable to compute higher order corrections of the dispersion law. Thus, we further collect $O(k)$ terms:
$$
\displaystyle
\tau_{xx}^{(1)}=2i\lambda\left((1-z)^{1-1/n}-(1-z)\right).
$$
\noindent
This function is integrable only for $n>1/2$. Next, we compute $p^{(1)}$:
\begin{equation}\label{eqp1l}
\displaystyle
\frac{R_e p^{(1)}}{F^2}=-\tau^{(1)}_{xx}-i\int_z^1 \tau_{xz}^{(0)}d\tilde z.
\end{equation}
\noindent
From this equation, we deduce that $\tau^{(2)}_{xz}$ is given by
$$
\begin{array}{ll}
\displaystyle
\tau^{(2)}_{xz}=-i\int_z^1\left(\frac{R_e p^{(1)}}{F^2}-\tau_{xx}^{(1)}+R_e(U_0'\phi^{(1)}-(\Lambda^{(0)}+U_0)(\phi^{(1)})'-\Lambda^{(1)}(\phi^{(0)})')\right)d\tilde z,\\
\displaystyle
\phi^{(2)}=-\int_0^z\int_0^{\tilde z}\frac{1}{n}(\lambda(1-y))^{1/n-1}\tau^{(2)}_{xz}+\phi^{(0)}dyd\tilde z.
\end{array}
$$
\noindent
We deduce that $\Lambda^{(2)}=\phi^{(2)}|_{z=1}$ is given by
{\setlength\arraycolsep{1pt}
\begin{eqnarray}
\displaystyle
\Lambda^{(2)}&=&R_e^2\left(\frac{4\,{\lambda}^{\frac{5}{n}-2}\,\left( 8\,{n}^{2}+25\,n+12\right) }{{\left( 2\,n+1\right) }^{2}\,\left( 3\,n+2\right) \,\left( 4\,n+3\right) \,\left( 5\,n+4\right) }\right)\nonumber\\
\displaystyle
&&-R_e{\cot}(\theta)\left(\frac{2\,{\lambda}^{\frac{3}{n}-1}\,\left( 5\,{n}^{2}+14\,n+6\right) }{{\left( 2\,n+1\right) }^{2}\,\left( 3\,n+2\right) \,\left( 4\,n+3\right) }
\right)\nonumber\\
\displaystyle
&&+\frac{4\,{\lambda}^{\frac{1}{n}}\,\left( n+2\right) }{3\,\left( 2\,n-1\right) \,\left( 3\,n+1\right) }
\end{eqnarray}}
\noindent
To the next order one finds:
$$
\begin{array}{ll}
\displaystyle
\tau^{(2)}_{xx}=-2i(\lambda(1-z))^{1-1/n}(\phi^{(1)})',\\
\displaystyle
\frac{R_e p^{(2)}}{F^2}=-\tau^{(2)}_{xx}-i\int_z^1\tau^{(1)}_{xz}d\tilde z-\int_z^1 R_e(U_0+\Lambda^{(0)})\phi^{(0)}.
\end{array}
$$
\noindent
This yields the following expressions for $(\tau^{(3)}_{xz},\phi^{(3)})$
$$
\begin{array}{ll}
\displaystyle
\tau^{(3)}_{xz}=-i\int_z^1\left(\frac{R_e p^{(2)}}{F^2}-\tau_{xx}^{(2)}+R_e(U_0'\phi^{(2)}-(\Lambda^{(0)}+U_0)(\phi^{(2)})'-\Lambda^{(1)}(\phi^{(1)})'-\Lambda^{(2)}(\phi^{(0)})')\right)d\tilde z,\\
\displaystyle
\phi^{(3)}=-\int_0^z\int_0^{\tilde z}\frac{1}{n}(\lambda(1-y))^{1/n-1}\tau^{(2)}_{xz}+\phi^{(1)}dyd\tilde z.
\end{array}
$$
\noindent
As a result, we deduce that $\Lambda^{(3)}=\phi^{(3)}|_{z=1}$ reads:
{\setlength\arraycolsep{1pt}
\begin{eqnarray}
\displaystyle
\Lambda^{(3)}&=&R_e^3\left(\frac{8\,i\,{\lambda}^{\frac{7}{n}-3}\,\left( 408\,{n}^{5}+3414\,{n}^{4}+8935\,{n}^{3}+9890\,{n}^{2}+4905\,n+900\right) }{{\left( 2\,n+1\right) }^{3}\,{\left( 3\,n+2\right) }^{2}\,\left( 4\,n+3\right) \,\left( 5\,n+4\right) \,\left( 6\,n+5\right) \,\left( 7\,n+6\right) }\right)\nonumber\\
\displaystyle
&&-R_e^2{\cot}(\theta)\left(\frac{3\,i\,{\lambda}^{\frac{5}{n}-2}\,\left( 244\,{n}^{5}+2248\,{n}^{4}+5675\,{n}^{3}+5940\,{n}^{2}+2776\,n+480\right) }{{\left( 2\,n+1\right) }^{3}\,{\left( 3\,n+2\right) }^{2}\,\left( 4\,n+3\right) \,\left( 5\,n+4\right) \,\left( 6\,n+5\right) }\right)\nonumber\\
\displaystyle
&&+R_e{\cot}^2(\theta)\left(\frac{2\,i\,{\lambda}^{\frac{3}{n}-1}}{{\left( 2\,n+1\right) }^{2}\,\left( 3\,n+2\right) }\right)-{\cot}(\theta)\left(\frac{i\,{\lambda}^{\frac{1}{n}}\,\left( 2\,n+7\right) }{3\,\left( 2\,n-1\right) \,\left( 4\,n+1\right) }\right)\nonumber\\
\displaystyle
&&-R_e\left(\frac{i\lambda^{\frac{3}{n}-1} \left( 988n^6+5666n^5+11622n^4+10825n^3+5008n^2+1120n+96     \right)        }{(2n-1)(2n+1)^2(3n+1)(3n+2)(4n+1)(5n+2)(5n+3)}\right)\nonumber
\end{eqnarray}}
\noindent
As a by product of this analysis, one recovers the classical values found in the expansion of the dispersion relation for a Newtonian fluid \cite{KK},\cite{B}($n=1$, $\lambda=3$ if the reference velocity is the mean value of the Nusselt solution):
{\setlength\arraycolsep{1pt}
\begin{eqnarray}
\displaystyle
\Lambda^{(0)}&=&-3,\nonumber\\
\displaystyle
\Lambda^{(1)}&=&i\left({\cot}(\theta)-\frac{6R_e}{5}\right),\nonumber\\
\displaystyle
\Lambda^{(2)}&=&-\frac{10}{7}\,R_e\,{\cot}(\theta)+\frac{12}{7}{R_e}^{2}+3,\nonumber\\
\displaystyle
\Lambda^{(3)}&=&i\left(\frac{75872}{25025}R_e^3-\frac{17363}{5775}R_e^2{\cot}(\theta)+\frac{2}{5}R_e{\cot}^2(\theta)+\frac{1413}{224}R_e{\cot}(\theta)-\frac{9}{5}{\cot}(\theta)\right).\nonumber
\end{eqnarray}}

\subsection{Comparison with other shallow water models}

The shallow water models proposed in this paper are consistent up to order $2$. In \cite{NM}, a first order shallow water model was proposed but also shown inconsistent since it did not predicted the correct stability threshold (derived from a consistent Benney type equation). Since then, other models was proposed in \cite{RQ}. In particular, in the shear thickening case $n>1$ where the regularization process does not introduce singularities, the authors proposed the shallow water model:
{\setlength\arraycolsep{1pt}
\begin{eqnarray}
\displaystyle
\partial_t h+\partial_x q&=&0,\nonumber\\
\displaystyle
R_e\partial_t q&=&R_e\left(-F(n)\frac{q}{h}\partial_x q+G(n)\frac{q^2}{h^2}\partial_x h\right)+I(n)\left(h(1-{\rm cotan}(\theta)\partial_x h)-\frac{q|q|^{n-1}}{(\phi_0h^2)^n}\right)\nonumber\\
\displaystyle
&&+(\frac{|q|}{\phi_0h^2})^{n-1}\left(J_1(n)\frac{q}{h^2}(\partial_x h)^2-K_1(n)\frac{\partial_x q\partial_x h}{h}-L_1(n)\frac{q}{h}\partial_{xx}h+M_1(n)\partial_{xx}q\right),\nonumber
\end{eqnarray}}
\noindent
with
$$
\begin{array}{ll}
\displaystyle
\phi_0=\frac{n}{2n+1},\quad F(n)=\frac{11n+6}{4n+3},\quad G(n)=\frac{6n+3}{4n+3},\quad I(n)=\frac{3n+2}{2(2n+1)},\\
\displaystyle
L_1(n)=\frac{(2n+1)(3n+2)(12n^3+36n^2+n-1)}{6n(2n-1)(3n+1)(4n+1)},\quad M_1(n)=\frac{(2n+1)(2n+7)(3n+7)}{6(4n+1)(2n-1)}.
\end{array}
$$
The coefficients $K_1(n)$ and $J_1(n)$ are unimportant here since we are going to study the stability of steady states under small wavenumber perturbations. For that purpose, we linearize the previous system about $(h,q)=(1,\phi_0)$. The system now reads
{\setlength\arraycolsep{1pt}
\begin{eqnarray}
\displaystyle
\partial_t h+\partial_x q&=&0,\nonumber\\
\displaystyle
R_e\partial_t q&=&R_e\left(G(n)\phi_0^2\partial_x h-F(n)\phi\partial_x q\right)\nonumber\\
\displaystyle
&&+I(n)\left((2n+1)(h-q)-{\rm cotan}(\theta)\partial_x h\right)\nonumber\\
\displaystyle
&&+M_1(n)\partial_{xx}q-L_1(n)\phi_0\partial_{xx}h.\nonumber
\end{eqnarray}
\noindent
We search for plane wave solutions in the form $(h,q)=e^{ik(\tilde\Lambda(k)\,t+x)}\left(\tilde h(k),\tilde q(k)\right)$ and expand both eigenvalues $\tilde\Lambda$ and $(\tilde h,\tilde q)$ with respect to $k\to 0$ as
$$
\displaystyle
(\tilde\Lambda,\tilde h,\tilde q)=\sum_{j=0}^2 k^j(\tilde\Lambda^{(j)},\tilde h^{(j)},\tilde q^{(j)})+O(k^3),
$$
\noindent
with the normalization $\tilde h^{(0)}=1$ and $\tilde h^{(j)}=0, j=1,2$. Inserting this expansion into the linearized system yields
{\setlength\arraycolsep{1pt}
\begin{eqnarray}
\displaystyle
\tilde\Lambda\tilde h+\tilde q&=&0,\nonumber\\
\displaystyle
iR_ek\tilde\Lambda\tilde q&=&ikR_e\left(G(n)\phi_0^2\tilde h-F(n)\phi_0\tilde q\right)\nonumber\\
\displaystyle
&&+I(n)\left((2n+1)(\tilde h-\tilde q)-ik{\rm cotan}(\theta)\tilde h\right)\nonumber\\
\displaystyle
&&+k^2\left(L_1(n)\phi_0\tilde h-M_1(n)\tilde q\right).\nonumber
\end{eqnarray}}
\noindent
By collecting $O(k^0)$ terms into this system, one obtains
$$
\displaystyle
\tilde\Lambda^{(0)}+\tilde q^{(0)}=0,\quad 1-\tilde q^{(0)}.
$$
\noindent
So that, one finds $\tilde\Lambda^{(0)}=-1$. Next, by collecting $O(k)$ terms into this system, one finds
$$
\displaystyle
\tilde\Lambda^{(1)}+q^{(1)}=0,\quad iR_e\tilde\Lambda^{(0)}\tilde q^{(0)}=iR_e\left(G(n)\phi_0^2-F(n)\phi_0\right)-I(n)\left((2n+1)\tilde q^{(1)}+i{\rm cotan}(\theta)\right).
$$
\noindent
From this system, we deduce
$$
\displaystyle
\tilde\Lambda^{(1)}=i\frac{{\rm cotan}(\theta)}{2n+1}-iR_e\frac{2}{(2n+1)(3n+2)}.
$$
\noindent
At this step, one can conclude that the shallow water system found in \cite{RQ} is consistent up to order one (at least at the linearized level).  At the second order, Ruyer Quil et al. pointed out that the terms quadratic in $\partial_x h,\partial_x q$ found as $n=1$ was not consistent with the order two shallow water model. Next, we show that even at the linearized level this system is not consistent except in the case $n=1$. Indeed, by collecting $O(k^2)$ terms in the linearized system yields
$$
\begin{array}{ll}
\displaystyle
\tilde\Lambda^{(2)}+\tilde q^{(2)}=0,\\
\displaystyle
iR_e(\Lambda^{(1)}\tilde q^{(0)}+\tilde\Lambda^{(0)}\tilde q^{(1)})=-iR_eF(n)\phi_0\tilde q^{(1)}-(2n+1)I(n)\tilde q^{(2)}+L_1(n)\phi_0-M_1(n).
\end{array}
$$
\noindent
From this equation, we find
{\setlength\arraycolsep{1pt}
\begin{eqnarray}
\displaystyle
\tilde{\Lambda}^{(2)}&=&R_e^2\frac{4\,\left( 5\,{n}^{2}+14\,n+6\right) }{{\left( 2\,n+1\right) }^{2}\,{\left( 3\,n+2\right) }^{2}\,\left( 4\,n+3\right) }\nonumber\\
\displaystyle
&&-R_e{\rm cotan}(\theta)\frac{2\,\left( 5\,{n}^{2}+14\,n+6\right) }{{\left( 2\,n+1\right) }^{2}\,\left( 3\,n+2\right) \,\left( 4\,n+3\right) }\nonumber\\
\displaystyle
&&+\frac{4\,\left( n+2\right) }{3\,\left( 2\,n-1\right) \,\left( 3\,n+1\right) }.\nonumber
\end{eqnarray}}
\noindent
One finds indeed that, in comparison to the $\Lambda^{(2)}$ of the Orr-Sommerfeld computations, 
$$
\displaystyle
\Lambda(k)|_{\lambda=1}-\tilde{\Lambda}(k)=R_e^2k^2\frac{4\,\left( 1-n\right) \,{n}^{2}}{{\left( 2\,n+1\right) }^{2}\,{\left( 3\,n+2\right) }^{2}\,\left( 4\,n+3\right) \,\left( 5\,n+4\right) }+O(k^3).
$$
\noindent
As a consequence, the model derived in \cite{RQ} for shear thickening fluids ($n>1$) is not consistent, even at the linear level with Navier-Stokes equations.: as quoted in \cite{RQ}, this is due to the fact that some corrective terms due to inertia are neglected.



\section{Conclusion and Perspectives}

In this paper, we have derived consistent shallow water equations for thin films of power law fluids down an incline. For that purpose, we have introduced a new, weaker, formulation of the Cauchy momentum equations. In particular, we have kept the fluid strain rate as an unknown in order to deal with the divergence of the apparent viscosity near the free surface. In this new framework, we have computed an expansion of the velocity field, in particular the bottom strain rate and streamwise discharge rate up to order 2 with respect to $\varepsilon$, the film ratio. We have then generated a {\it family} of consistent models of shallow water type. Up to first order, we proposed a new model that is consistent, conservative and invariant by Galilean transformation. Up to second order, we proposed the first fully consistent shallow water model that is an extension of the one proposed in \cite{RQ}.

   One classical test to verify the consistency of shallow water models is to compare the spectrum associated to the linearized equations about a constant state and the one that arises from a direct analysis of the Cauchy momentum equations. As a byproduct of our new formulation of Cauchy momentum equations, we propose the first rigorous spectral analysis of power law fluids down an incline. We have derived a suitable linear problem that is an generalization of Orr Sommerfeld equations for newtonian fluids. We have shown that the nonlinear expansion of the velocity field in the shallow water scaling provides a natural expansion of eigenvalues and eigenfunctions of this generalized Orr Sommerfeld equations.
   
Our analysis is restricted to $n>1/2$ and strongly suggests that the free surface Cauchy momentum equations are ill posed if $n<1/2$ even at the linearized level. As a consequence, the use of the power law in some applications like glaciology $n=1/3$ or for various polymers is questionable: in such situations, one has to {\it regularize} the power law at low strain rate. In \cite{RQ}, the authors used such a regularization procedure in order to compute shallow water models for shear thinning fluids $n<1$ whereas it is useless if $n>1$. Here, we show that both cases are treated similarly provided that $n>1/2$: as a result, one {\it has to} recover our models and expansions when the parameter of regularization goes to zero if $n>1/2$. In contrast, if $n<1/2$, one needs a regularization procedure to derive shallow water models and these models should strongly depend on the regularization procedure: this question shall be treated soon in a future work. 

There are various exemples of regularization of the power law found in the literature like the Carreau law or mixed power law/newtonian law used in \cite{RQ}. In the later case, a strain threshold is introduced: below this threshold, the fluid is newtonian and above, it is a power law. The authors introduced this law in order to do computations explicitly but the threshold is fixed which may leads to some inconsistency if one let this threshold goes to zero. The Carreau law seems more natural but computations are not explicit. Then, a generalization of this law that would lead to explicit computations is of interest in order to compare with our results: this problem and the influence on the derivation of shallow water models, stability properties of constant states are very interesting for applications purposes (sea of ice, flows of polymers with $n<1/2$). Another interesting extension to this work would be to study the influence of the boundary conditions at the bottom, in particular if the no-slip condition is replaced by the Navier-slip condition: this is of interest if one considers flows of power law fluids down porous media \cite{UMBR}

\end{document}